\documentclass[10pt, Conference, letterpaper]{IEEEtran}
\IEEEoverridecommandlockouts
\usepackage{cite}
\usepackage{booktabs}
\usepackage[T1]{fontenc}
\usepackage{aecompl}
\usepackage{subcaption}
\usepackage[numbers,sort&compress]{natbib}
\usepackage{amsmath,amssymb,amsfonts}
\usepackage{amsthm}
\usepackage{float}
\usepackage{booktabs}
\makeatletter
\renewcommand{\maketag@@@}[1]{\hbox{\m@th\normalsize\normalfont#1}}%
\makeatother
\usepackage{graphicx}
\usepackage{verbatim}
\usepackage{textcomp}
\usepackage{xcolor}
\usepackage{stfloats}
\usepackage{tabularx}
\usepackage{array}
\usepackage{makecell}
\usepackage{url}
\makeatother
\usepackage{pifont}
\usepackage{textcomp}
\usepackage[marginal]{footmisc}
\usepackage{geometry}
\usepackage[hidelinks]{hyperref}
\usepackage{setspace}
\usepackage{caption}
\allowdisplaybreaks[4]
\usepackage{algpseudocode}
\usepackage[ruled,linesnumbered]{algorithm2e}

\def\BibTeX{{\rm B\kern-.05em{\sc i\kern-.025em b}\kern-.08em
    T\kern-.1667em\lower.7ex\hbox{E}\kern-.125emX}}
\newtheorem{myDef}{Definition}

\begin{document}
\title{Topology-aware Microservice Architecture in Edge Networks: Deployment Optimization and Implementation}
\author{Yuang Chen, \IEEEmembership{Graduate Student Member, IEEE}, Chang Wu, Fangyu Zhang, Chengdi Lu, Yongsheng Huang, and Hancheng Lu, \IEEEmembership{Senior Member, IEEE} \thanks{\setlength{\baselineskip}{1.5\baselineskip} Yuang Chen, Chang Wu, Fangyu Zhang, Chengdi Lu, Yongsheng Huang, and Hancheng Lu are with the University of Science and Technology of China, Hefei, 230027, China. (email: $\{$yuangchen21, changwu, fv215b, lcd1999, ysh6$\}$@mail.ustc.edu.cn; hclu@ustc.edu.cn). Hancheng Lu is also with the Institute of Artificial Intelligence, Hefei Comprehensive National Science Center, Hefei, 230088, China.}}

\maketitle
\begin{abstract}
As a ubiquitous deployment paradigm, integrating microservice architecture (MSA) into edge networks promises to enhance the flexibility and scalability of services. However, it also presents significant challenges stemming from dispersed node locations and intricate network topologies. In this paper, we have proposed a topology-aware MSA characterized by a three-tier network traffic model encompassing the service, microservices, and edge node layers. This model meticulously characterizes the complex dependencies between edge network topologies and microservices, mapping microservice deployment onto link traffic to accurately estimate communication delay. Building upon this model, we have formulated a weighted sum communication delay optimization problem considering different types of services. Then, a novel topology-aware and individual-adaptive microservices deployment (TAIA-MD) scheme is proposed to solve the problem efficiently, which accurately senses the network topology and incorporates an individual-adaptive mechanism in a genetic algorithm to accelerate the convergence and avoid local optima. Extensive simulations show that, compared to the existing deployment schemes, TAIA-MD improves the communication delay performance by approximately 30\% to 60\% and effectively enhances the overall network performance. Furthermore, we implement the TAIA-MD scheme on a practical microservice physical platform. The experimental results demonstrate that TAIA-MD achieves superior robustness in withstanding link failures and network fluctuations.
\end{abstract}

\vspace{-0.6em}

\section{Introduction}
\par With the proliferation of smartphones and internet-of-things (IoT) devices \cite{chen2024performance}, along with the widespread adoption of cloud and edge computing and the explosive growth of digital content and services, the scale of users and the resource demands for network services have surged dramatically \cite{hannousse2021securing}. For instance, ChatGPT reached over one billion monthly active users within less than two months of its launch \cite{10113601}. Currently, mobile applications are increasingly focusing on being location-aware, compute-intensive, and latency-sensitive \cite{chen2024performance,10382447,10355071}. Combined with continuously growing user bases and increasing quality-of-service (QoS) requirements, the limitations of cloud computing, which centralizes service operations in the cloud, are gradually becoming apparent \cite{sandhu2021big,jauro2020deep}. Edge computing has emerged as an advanced computing paradigm that enables the deployment of computing and storage resources closer to users, significantly improving real-time performance and reducing data traffic costs \cite{10430407,10460318,10529607,elbamby2019wireless}. To fully utilize computing resources, edge computing must reallocate hardware and software resources via virtualization technologies to serve multiple users on the same hardware \cite{9063490}. Nevertheless, compared to centralized cloud networks, edge networks face tougher challenges in terms of user mobility, device heterogeneity, limited resources, and geographically dispersed edge nodes \cite{pallewatta2023placement,10128791,mansouri2021review}. Moreover, commonly used virtualization technologies based on virtual machines (VMs) are impractical for edge environments due to their high resource overhead, slow startup times, and complex deployment and migration processes \cite{xu2013managing,8967018}.

\par To alleviate the above issues, a more distributed and service-oriented application architecture \cite{pahl2016microservices, 8951173}, microservice architecture (MSA), has recently emerged as a potential enabler for flexible service deployment in distributed scenarios \cite{9615028,9774016,9154603,10128791,gu2022layer,zeng2023layered}. In MSA, complex monolithic applications are decomposed into multiple logically distinct, mutually complementary, and single-function microservices, which cooperate with each other to provide combined functionality \cite{10128791,9057418,8951173}. This approach offers significant flexibility, robustness, and scalability for the deployment of distributed services. Each microservice in MSA can be independently developed, upgraded, and deployed using different programming languages, thus effectively addressing the scalability, maintenance, and team collaboration issues inherent in traditional monolithic applications \cite{wang2021promises}. Currently, many internet giants such as Amazon, Netflix, and eBay have extensively exploited MSA, and their research and development progress reports claim the effectiveness of MSA \cite{aksakalli2021deployment}. For instance, Amazon emphasized in \cite{aws_microservice_extractor} that MSA can achieve more controllable and faster production, easily expand computing resources, reduce unit complexity, and create more scalable development teams. In addition, microservices are typically deployed using lightweight virtualization technologies like \emph{containers} \cite{10098822}, which leads to faster deployment and startup, lower resource consumption, and greater compatibility, and can be flexibly deployed on resource-constrained, platform-heterogeneous edge or fog devices \cite{pallewatta2022qos}. Consequently, developing MSA-based service deployment approaches for edge networks is expected to enhance the flexibility and scalability of services, becoming one of the mainstream technologies for service deployment.

\vspace{-0.8em}

\subsection{Motivations and Challenges}

\begin{figure}[t]
\vspace{-0.5em}
\centering
\includegraphics[scale=0.4]{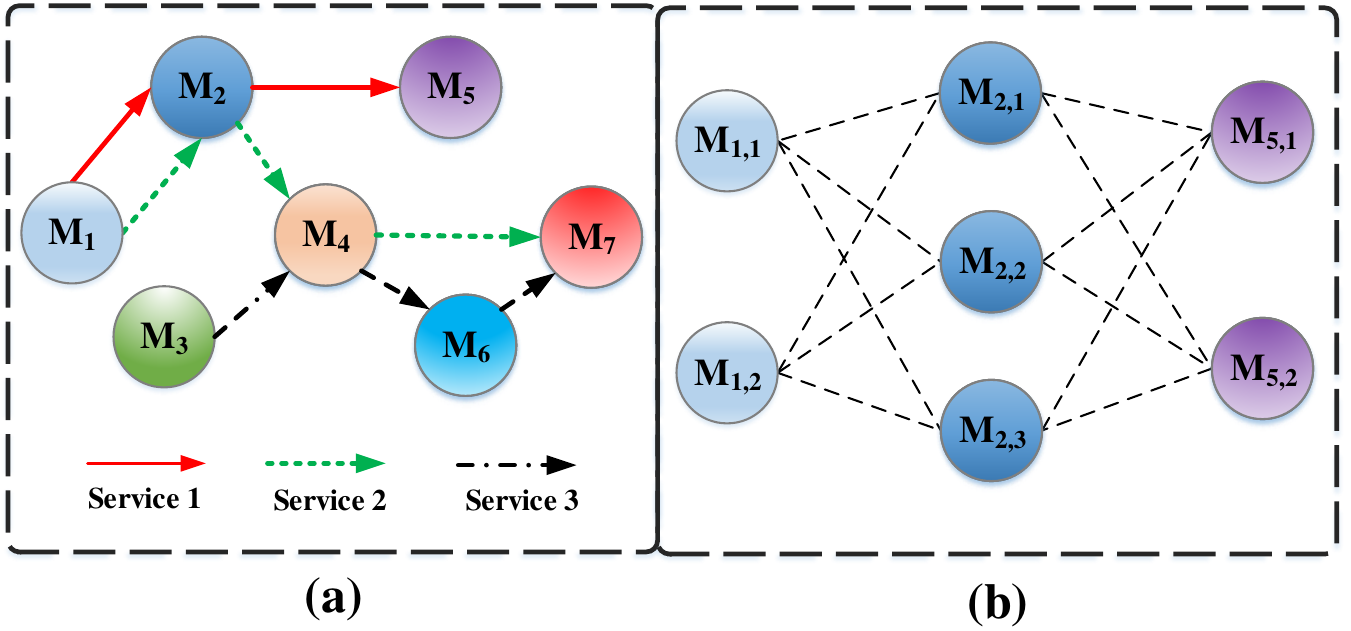}
\vspace{-1em}
\caption{\small Complex dependencies among microservices. (a) Microservice invocations for services. (b) Invocations among microservice instances.}
\label{fig1}
\vspace{-1.0em}
\end{figure}

\begin{figure}[t]
\centering
\includegraphics[scale=0.27]{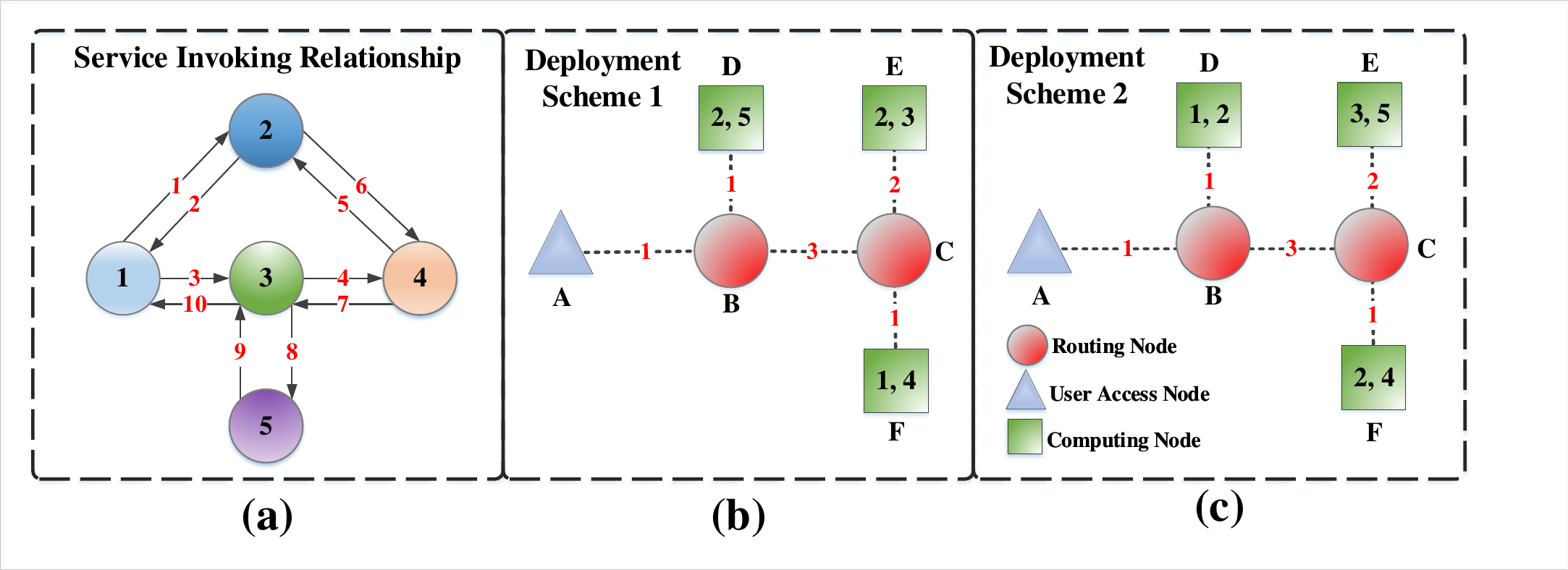}
\vspace{-1em}
\caption{\small The impacts of microservice deployment schemes on the QoS of services. (a) Microservice Invocations. (b) Deployment Scheme 1. (c) Deployment Scheme 2.}
\label{fig2}
\vspace{-1.8em}
\end{figure}

\par Although MSA is anticipated to provide finer-grained deployment schemes for edge networks, enhancing flexibility and scalability, and speeding up application development cycles, the development of microservices in edge networks still encounters numerous stringent challenges.

\subsubsection{Complex microservice dependencies in edge networks with geographically dispersed nodes} The geographical dispersion of edge nodes necessitates complex topology forwarding for inter-device communications, which is exacerbated by the limited and unstable link bandwidth, resulting in significant differences in communication delay and bandwidth occupancy across different nodes \cite{pallewatta2023placement,10128791}. In addition, microservices in MSA typically have complex dependencies \cite{9740415}, as illustrated in Fig. \ref{fig1}, where $M_{i}$ represents a type of microservice and $M_{i,j}$ denotes the $j$-th instance of $M_{i}$. Fig. \ref{fig1} (a) shows that fulfilling a service requirement usually necessitates the collaboration of multiple microservices, with various services potentially sharing the same type of microservice. As shown in Fig. \ref{fig1} (b), each type of microservice can have multiple instances with complex invocation relationships among them. Consequently, compared to traditional monolithic applications, MSA demands greater communication requirements and exhibits more intricate inter-instance relationships. This complicates traffic analysis between edge nodes, and existing work lacks effective analysis models \cite{10162207, 9154603, 9615028}. In this regard, an effective topology-aware microservice deployment model is necessitated to map microservice deployments onto edge nodes' link traffic, thereby accurately characterizing the complex edge network topology as well as inter-microservice dependencies.

\subsubsection{Difficulties of implementing optimal microservice deployment in edge networks} The essence of microservice deployment involves shared resource allocation and communication overhead, which directly impacts the QoS of microservice-based applications \cite{9615028}. As illustrated in Fig. \ref{fig2}, the impacts of microservice deployment schemes on the QoS of services are presented. In Fig. \ref{fig2} (a), the microservice invocations required for a specific service are depicted, where the node labels represent the microservice types and the arrow labels indicate the order of execution for the request or response. Fig. \ref{fig2} (b) and Fig. \ref{fig2} (c) show two different deployment schemes for the microservice invocations described in Fig. \ref{fig2} (a), where node labels represent the deployed microservice types and the connecting line labels denote the inter-node communication delay (in \emph{ms}). Compared to deployment scheme 1, deployment scheme 2 significantly reduces communication delay by collocating frequently interacting microservice instances on the same or nearby computing nodes. In particular, based on the microservice invocations depicted in Fig. \ref{fig2} (a), from the moment a user request is received at the user-access node $\mathbf{A}$ to the moment a user response returns to node $\mathbf{A}$, the total communication delays generated by deployment schemes 1 and 2 are $46$ \emph{ms} and $22$ \emph{ms}, respectively. Moreover, deployment scheme 2 deploys instances of microservice $\mathbf{2}$ on both computing nodes $\mathbf{D}$ and $\mathbf{F}$, which allows microservice $\mathbf{1}$ on computing node $\mathbf{D}$ to invoke microservice $\mathbf{2}$ locally, thereby avoiding cross-node communication and reducing communication delay and overhead \cite{9042873}.

\par The aforementioned analysis highlights that a well-designed microservice deployment scheme in distributed edge networks can remarkably conserve network resources and reduce communication delay. Due to the limited computing capacity of edge nodes, multiple microservice instances are usually deployed across distributed nodes to meet the QoS requirements of applications and avoid single points of failure \cite{9669179, 9042873}, which can effectively combat cross-node communication and further reduce communication delay and overhead \cite{9042873}. What's more, deploying multiple instances on the same edge node may lead to resource contention, and allocating more resources to one microservice may slow down the operation speed of other microservices \cite{9615028}. As a result, microservice deployment optimization must be dynamically adjusted according to service load since microservices compete for resources with each other, and the QoS of the service is affected by almost all microservices.

\begin{figure}[t]
\vspace{-0.5em}
\centering
\includegraphics[scale=0.55]{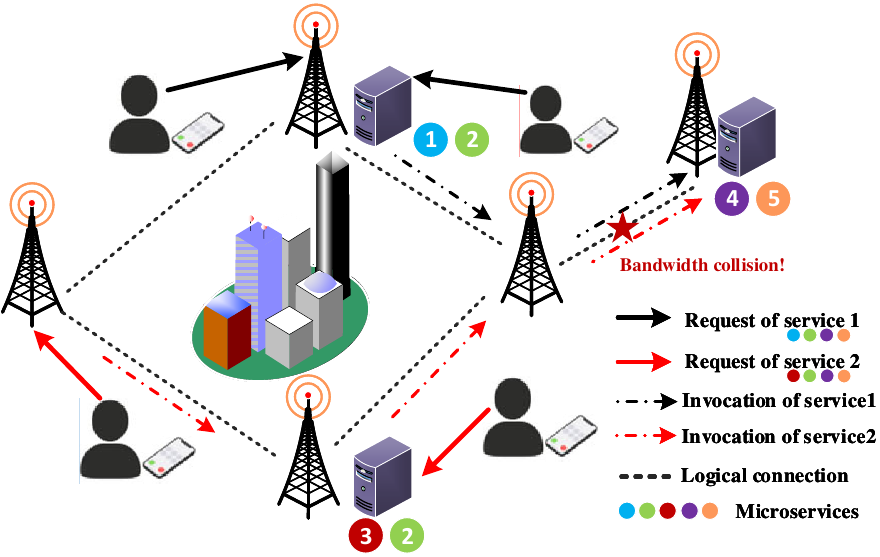}
\vspace{-0.6em}
\caption{\small An example of bandwidth collision in edge networks.}
\label{fig3}
\vspace{-2em}
\end{figure}

\subsubsection{The prerequisite of optimal microservice deployment necessitates establishing an effective traffic analysis model} In edge networks with massive services and users, some links carry multiple service traffic concurrently, leading to competition for network resources \cite{9615028}. Fig. \ref{fig3} illustrates the bandwidth collisions of service traffic in edge networks, where two types of services are realized through mutual invocation among various microservices. Both \textbf{service 1} and \textbf{service 2} need to invoke \textbf{microservice 4} and \textbf{microservice 5}, resulting in bandwidth collisions as their traffic inevitably crosses the link marked with the red star in Fig. \ref{fig3}. Therefore, optimizing microservice deployment necessitates an effective traffic analysis model to accurately capture complex microservice dependencies and network topologies. However, this is extremely difficult. Existing studies usually simplify edge networks by considering the bandwidth and delay between nodes as fixed values, which fails to fully incorporate network topology and ignores the impact of deployment schemes on link bandwidth, resulting in lower-than-expected communication capacity between instances and even congestion \cite{9740415, 9615028, 9869329}. Furthermore, most studies focus only on simulation validation and lack actual physical platform verification, which ignores the feasibility of microservice deployment schemes deployed in real systems and leads to inaccurate estimation of link bandwidth. Therefore, implementing and validating the performance of deployment schemes in actual physical validation platforms is essential for optimal microservice deployment schemes in edge networks with intricate topologies \cite{bugshan2023intrusion,pallewatta2023placement,gu2022layer,zeng2023layered}.

\vspace{-1em}

\subsection{Main Contributions}

\par To effectively overcome the aforementioned challenges, we have proposed an innovative topology-aware MSA, which incorporates a three-tier network traffic analysis model. Then, we have formulated a weighted sum communication delay optimization problem that aims to improve delay performance by optimizing microservices deployment. To effectively tackle this problem, we have developed a novel topology-aware and individual-adaptive microservices deployment (TAIA-MD) scheme. Extensive simulations and physical platform validations demonstrate the superior performance of our proposed TAIA-MD scheme. The primary contributions of this paper are summarized as follows:

\begin{itemize}
  \item For the first time, we have proposed an innovative topology-aware MSA, which features a three-tier network traffic analysis model comprising the service, microservice, and edge node layers. This model meticulously captures the complex inter-microservice dependencies and edge network topologies by mapping microservice deployments onto edge nodes' link traffic to accurately estimate the communication delay.

  \item Building upon this model, we have formulated a weighted sum communication delay optimization problem that considers the load conditions of services, aiming to improve delay performance through optimizing microservices deployment. To address this intractable problem, we have developed a novel microservice deployment scheme named TAIA-MD, which customizes the deployment scheme by sensing network topologies and incorporates an individual-adaptive mechanism in the genetic algorithm (GA) to accelerate the convergence and avoid local optima.

  \item Extensive simulations demonstrate that TAIA-MD can reduce communication delay by approximately 30\% to 60\% compared to existing deployment schemes in bandwidth-constrained and topology-complex edge networks. Moreover, physical platform experiments further show the superior robustness of the TAIA-MD in effectively combating link failures and network fluctuations.
\end{itemize}

\par The remainder of this paper is organized as follows. In Section II, we provide an in-depth literature review of relevant studies. The topology-aware MSA is introduced in Section III. Section IV formulates and solves the weighted sum communication delay optimization problem. The results and analysis of simulations and physical verifications are provided in Section V. Finally, Section VI concludes this paper and discusses future directions.

\vspace{-1em}

\section{RELATED WORKS}
\par In this section, we first review earlier literature primarily focused on microservice deployments centered around cloud environments. Then, we comprehensively summarize and examine the latest microservice deployment schemes in edge networks. Finally, we identify limitations related to the aforementioned challenges from these literature reviews.

\vspace{-1.5em}

\subsection{Deployment Schemes Centered around Cloud Environments}

\par Due to the highly decoupled nature of MSA, implementing a certain service usually requires close collaboration among numerous microservices \cite{heinrich2017performance,10418890}. However, deploying hundreds or thousands of microservice instances on resource-constrained servers presents a formidable challenge \cite{10592806,10418890,pallewatta2023placement,gu2022layer,zeng2023layered}. Research on microservice deployment schemes is always of paramount importance. Early studies on microservice deployment primarily centered around cloud environments \cite{10013701,10024362,10589861}. In \cite{10013701}, the authors have considered resource contention and deployment time among microservices, and proposed a parallel deployment scheme to aggregate microservices that compete for as diverse resources as possible, thereby minimizing interference in resource-constrained cloud systems. To alleviate the enormous pressure brought by the scale of microservices on cluster management in cloud computing, the authors in \cite{10024362} have developed a topology-aware deployment framework that leverages a heuristic graph mapping algorithm to capture the topological structure of microservices and clusters, thereby minimizing network overhead for applications. In order to mitigate communication delays caused by intricately interdependent microservices, the authors in \cite{10589861} have explored a novel deployment scheme that leverages the fine-grained and comprehensive modeling of microservice dependencies.

\vspace{-1.3em}

\subsection{Latest Deployment Schemes for Edge Networks}
\par Microservice deployment in edge networks is receiving increasing attention due to the flourishing development of latency-sensitive services, as well as advancements in edge or fog computing techniques. In \cite{9615028} and \cite{9460542}, the authors have proposed a novel MSA called Nautilus to deploy microservice instances in the cloud-edge continuum effectively. This MSA adjusts instance placement and achieves load balancing by migrating containers from busy nodes, thereby ensuring the required QoS under external shared resource contention. To mitigate the impact of network load and routing on service reliability, in \cite{Fangyu_Zhang}, we have proposed a network-aware service reliability model that effectively captures the correlation between network state changes and reliability. In addition, we have designed a service reliability-aware deployment algorithm, which leverages shared path reliability calculation to improve service failure tolerance and bandwidth consumption. In \cite{9993766}, the authors have studied an optimal microservice deployment scheme to balance layer sharing and chain sharing in resource-constrained edge servers. This scheme effectively addresses the challenges of microservice image pull delay and communication overhead minimization.

\vspace{-1em}

\subsection{Limitations of Literature Reviews}

\par Although current research reveals valuable insights into microservice deployment in edge networks, it seldom accounts for the tangible impact of edge environments and resource contention \cite{heinrich2017performance,10418890,10592806,pallewatta2023placement,gu2022layer,zeng2023layered,10013701,10024362,10589861,9615028,9460542,Fangyu_Zhang,9993766}. The widespread distribution of edge devices and nodes, multi-hop forwarding characteristics of inter-node communications, and the complex intrinsic dependencies among microservices make accurate traffic analysis modeling a mystery \cite{9740415, 9615028, 9869329}. Concurrently, the coexistence of multiple flows on numerous links sparks resource competition, leading to significant variations in link performance across deployment strategies. Moreover, improper deployment might overwhelm links, drastically degrading data transfer rates and increasing service delays. Nevertheless, prevailing studies typically simplify traffic analysis by disregarding the dynamic nature of network topology, node bandwidth, and delay \cite{9869329}.

\vspace{-0.5em}

\section{Topology-aware Microservice Architecture for Edge Networks}

\begin{figure}[t]
\vspace{-0.5em}
\centering
\includegraphics[scale=0.3]{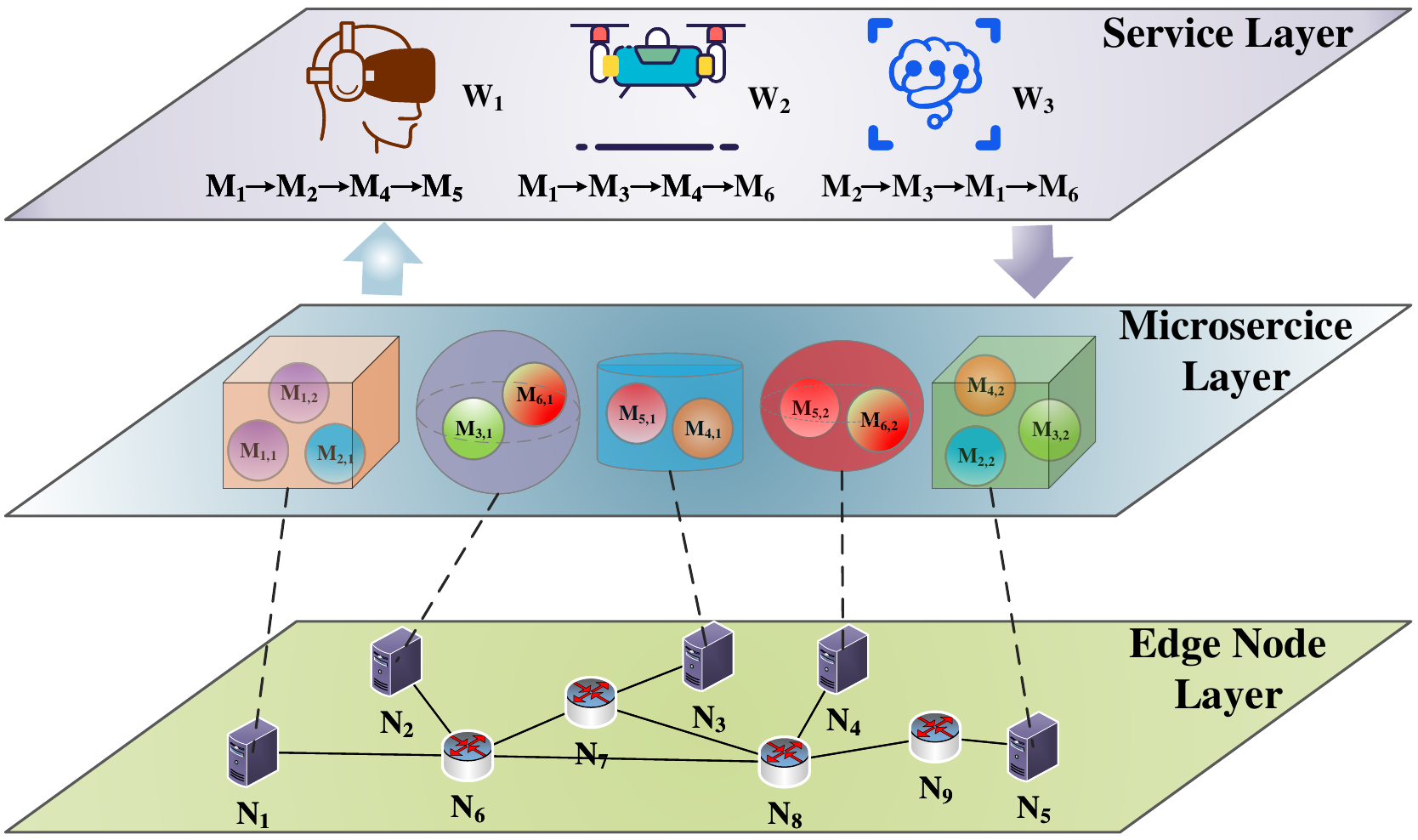}
\vspace{-0.6em}
\caption{\small Topology-aware microservice deployment architecture.}
\label{fig4}
\vspace{-2em}
\end{figure}

\par As illustrated in Fig. \ref{fig4}, we propose an innovative topology-aware microservice architecture characterized by a three-tier network traffic analysis model, including service layer invocation model, microservice layer data model, and edge-node layer traffic model. This model maps microservice deployments onto edge nodes' link traffic to meticulously capture the complex inter-microservice dependencies and edge network topology. Let $W_{k}$ and $N_{p}$ indicate the $k$-th type of service and the $p$-th node, respectively, where $k \! \in \! \mathcal{K} \! \triangleq \! \left\{1,\cdots, K\right\}$ and $p \! \in \! \mathcal{P} \! \triangleq \! \left\{1,\cdots, P\right\}$. The $i$-th type of microservice is denoted as $M_{i}$, where $i \! \in \! \mathcal{I} \! \triangleq \! \left\{1, \cdots, I\right\}$. Let $M_{i,a}$ denote the $a$-th instance of $M_{i}$, where $a \! \in \! \mathcal{A}_{i} \
\! \triangleq \! \left\{1,\cdots,A_{i}\right\}$ and $\mathcal{A}_{i}$ is the set of instances for $M_{i}$.

\par For example, consider microservice instances deployed across five computing nodes $N_{1} \!\! \sim \!\! N_{5}$. Notably, our proposed microservice architecture is not limited to this example. Instead, it is designed with broad applicability. As shown in Fig. \ref{fig4}, when $W_{1}$ is requested, microservices will be invoked in the order $M_{1} \! \rightarrow \! M_{2} \! \rightarrow \! M_{4} \! \rightarrow \! M_{5}$. One possible sequence of instance invocations at the microservice layer can be denoted as $M_{1,1} \!\! \rightarrow \!\! M_{2,1} \!\! \rightarrow \!\! M_{4,1} \!\! \rightarrow \!\! M_{5,2}$, which are deployed on nodes $N_{1}$, $N_{3}$, and $N_{4}$. In this case, the core of minimizing communication delay in service execution at the edge-node layer rests with how to optimize the deployment locations of microservices in the microservice layer.

\vspace{-0.8em}

\subsection{Service Layer Invocation Model}
\par
\begin{figure}[h]
\vspace{-0.5em}
\centering
\includegraphics[scale=0.4]{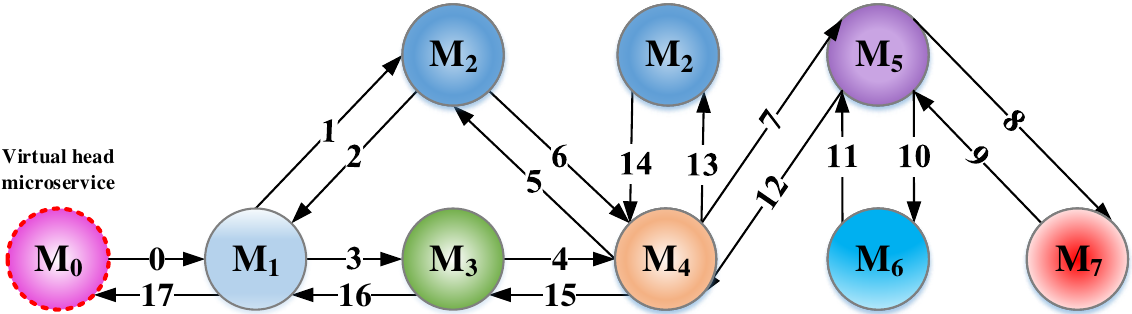}
\vspace{-0.3em}
\caption{A particular service invocation based on virtual head microservice.}
\label{fig5}
\vspace{-1.2em}
\end{figure}

 First of all, we introduce the service layer invocation model. To facilitate clarity, we illustrate it with a specific example. As shown in Fig. \ref{fig5}, we examine a particular service invocation executing the service $W_{k}, k \! \in \! \mathcal{K}$. The labels inside each node indicate the type of microservice, while the labels on each arrow represent the execution order of requests or responses. In edge networks, users access the network through their nearby user-access nodes. In particular, user's requests first arrive at the user-access nodes and are then forwarded to the appropriate computing nodes based on the deployment scheme. To describe the process of user requests transitioning from the user-access node to the first microservice, we introduce a special microservice type $M_{0}$ called the virtual head microservice that is occupancy-free of any computational resources.

\par Let $F_{i,j}$ and $R_{i,j}, i, j \in \mathcal{I}, i \neq j$ indicate a request and response from $M_{i}$ to $M_{j}$ during the execution of service $W_{k}$, respectively. In this case, the microservice invocation of service $W_{k}$ presented in Fig. \ref{fig5} can be formulated using multiset as follows:

\vspace{-0.7em}

\begin{equation}\label{eq1}
   \begin{aligned}
      W_{k} \Longrightarrow & \bigg\{F_{0,1}, F_{1,2}, R_{2,1}, F_{1,3}, F_{3,4}, F_{4,2}, R_{2,4}, F_{4,5}, F_{5,7},\\
                            & R_{7,5}, F_{5,6}, R_{6,5}, R_{5,4}, F_{4,2}, R_{2,4}, R_{4,3}, R_{3,1}, R_{1,0}\bigg\} \\
            \Longrightarrow & \bigg\{F_{0,1}, F_{1,2}, F_{1,3}, F_{3,4}, 2F_{4,2}, F_{4,5}, F_{5,6}, F_{5,7},\\
                            & R_{2,1}, 2R_{2,4}, R_{3,1}, R_{4,3}, R_{5,4}, R_{6,5}, R_{7,5}, R_{1,0}\bigg\}.
   \end{aligned}
\end{equation}

\par We define $\mathcal{F}_{k}$ as the request matrix for $W_{k}$, where $F_{i,j}^{k} \in \mathcal{F}_{k}$ represent the number of requests from $M_{i}$ to $M_{j}$ during the execution of service $W_{k}$. Thus, $\mathcal{F}_{k}$ can be given by

\begin{small}
\begin{equation}\label{eq2}
    \mathcal{F}_{k} = \begin{bmatrix}
        0 & 1 & 0 & 0 & 0 & 0 & 0 & 0 \\
        0 & 0 & 1 & 1 & 0 & 0 & 0 & 0 \\
        0 & 0 & 0 & 0 & 0 & 0 & 0 & 0 \\
        0 & 0 & 0 & 0 & 1 & 0 & 0 & 0 \\
        0 & 0 & 2 & 0 & 0 & 1 & 0 & 0 \\
        0 & 0 & 0 & 0 & 0 & 0 & 1 & 1 \\
        0 & 0 & 0 & 0 & 0 & 0 & 0 & 0 \\
        0 & 0 & 0 & 0 & 0 & 0 & 0 & 0 \\
    \end{bmatrix}
    .
\end{equation}
\end{small}

\par Similarly, the response matrix of service $W_{k}$ can be defined by $\mathcal{R}^{k}=\left(\mathcal{F}_{k}\right)^{\mathrm{T}}$.

\vspace{-1em}

\subsection{Microservice Layer Data Model}
\par Each type of microservice is usually equipped with multiple instances, and this paper considers scheduling them using the Round Robin method. Thus, the likelihood of each instance $M_{i, a}$ for microservice $M_{i}$ being invoked is the same and can be given as follows:

\vspace{-0.3em}

\begin{equation}\label{eq3}
   P_{M_{i,a}} = \frac{1}{\left|\mathcal{A}_{i}\right|}, \forall i \in \mathcal{I}\backslash \left\{0\right\}, a \in \mathcal{A}_{i}.
\end{equation}

\par For the case of $i = 0$, it depends on the service arrival rate at each user-access node, which will be explained in detail later.

\par Let $P_{M_{i,a}}^{k}$ denote the probability of selecting instance $M_{i,a}$ when invoking microservice $M_{i}$ during the execution of service $W_{k}$, where $i \in \mathcal{I}, k \in \mathcal{K}, a \in \mathcal{A}_{i}$. Let $F_{M_{i,a}, M_{j,b}}^{k}$ represent the average frequencies instance $M_{i,a}$ requests instance $M_{j,b}$ during the execution of service $W_{k}$, then we have

\vspace{-0.7em}

\begin{equation}\label{eq4}
   F_{M_{i,a}, M_{j,b}}^{k} = F_{i,j}^{k} P_{M_{i,a}}^{k} P_{M_{j,b}}^{k}, a \in \mathcal{A}_{i}, b \in \mathcal{A}_{j}, i \neq j.
\end{equation}

\par Let $D_{i,k}^{req}$ indicate the average requested data size when $M_{i}$ invokes $M_{j}$, and $\widetilde{F}_{M_{i,a}, M_{j,b}}^{k}$ denote the average data size requested by instance $M_{i,a}$ from instance $M_{j,b}$ during the execution of service $W_{k}$. Then we can obtain that

\vspace{-0.5em}

\begin{equation}\label{eq5}
   \widetilde{F}_{M_{i,a}, M_{j,b}}^{k} = D_{i,k}^{req} F_{M_{i,a}, M_{j,b}}^{k}.
\end{equation}

\par Similarly, let $R_{M_{i,a}, M_{j,b}}^{k}$ indicate the average frequencies that the instance $M_{i,a}$ responds to instance $M_{j,b}$ when $W_{k}$ is invoked, where $i,j \in \mathcal{I}, a \in \mathcal{A}_{i}, b \in \mathcal{A}_{j}$. The average data size per invocation from $M_{i}$ to $M_{j}$ and from $M_{i,a}$ to $M_{j,b}$ can be expressed as $D_{i,j}^{res}$ and $\widetilde{R}_{M_{i,a}, M_{j,b}}^{k}$, respectively. Since the request and response processes correspond to each other, we can obtain that

\vspace{-1em}

\begin{subequations}\label{eq6}
   \begin{align}
      & {R}_{M_{i,a},M_{j,b}}^{k}={F}_{M_{j,b},M_{i,a}}^{k},\\
      & \widetilde{{R}}_{M_{i,a},M_{j,b}}^{k}=D^{res}_{i,j}{R}_{M_{i,a},M_{j,b}}^{k}.
   \end{align}
\end{subequations}

\vspace{-0.5em}

\subsection{Edge-Node Layer Traffic Model}
\par The edge-node layer comprises three types of nodes: user-access nodes for processing user requests, routing nodes for data forwarding, and computing nodes for deploying and executing microservices. All nodes can be represented as set $\mathcal{N}$, with user-acess nodes as set $\mathcal{N}_{acc}$, routing nodes as set $\mathcal{N}_{rou}$, and computing nodes as set $\mathcal{N}_{cmp}$. In this case, we have $\mathcal{N} \triangleq \mathcal{N}_{acc} \cup \mathcal{N}_{rou} \cup \mathcal{N}_{cmp}$. We consider service requests arriving at each user-access node follow a Poisson distribution. If the Poisson parameter for $W_{k}$ at user-access node $N_{p} \in \mathcal{N}_{acc}$ is $\lambda_{p}^{k}$, then the Poisson parameter for the request of service $W_{k}$ in the edge-node layer can be expressed as

\vspace{-0.5em}

\begin{equation}\label{eq7}
   \lambda^{k} = \sum\limits_{N_{p} \in \mathcal{N}_{acc}} \lambda_{p}^{k}.
\end{equation}

\vspace{-0.5em}

\par The computing node for deploying instance $M_{i,a}$ can be represented as $L_{i,a} \in \mathcal{N}_{cmp}$. For instances of the virtual head microservice $M_{0}$, the invocation probability of invoking instance $M_{0,c}$ during the execution of service $W_{k}$ can be given by

\vspace{-0.5em}

\begin{equation}\label{eq8}
   P_{M_{0,c}}^{k} = \frac{\lambda_{L_{0,c}}^{k}}{\lambda^{k}}, c \in \mathcal{A}_{0}.
\end{equation}

\par In this case, once microservice $M_{i}$ is invoked during the execution of $W_{k}$, the invocation probability of the instance $M_{i,a}$ can be denoted as follows:

\begin{equation}\label{eq9}
    P_{M_{i,a}}^{k} =\begin{cases}
        \frac{\lambda^{k}_{L_{0,a}}}{\lambda^{k}}, & \text{ if } i = 0, a \in \mathcal{A}_{0},\\
        \frac {1}{\left |\mathcal{A}_{i}\right | },  & \text{ if } i \neq 0, a \in \mathcal{A}_{i}.
    \end{cases}
\end{equation}

\par Given any two nodes $N_{p}$, $N_{q} \in \mathcal{N}$, we define the traffic that directly flows from $N_{p}$ to $N_{q}$ without any forwarding as the direct traffic $S_{p,q}$. If nodes $N_{x}$ and $N_{y}$ are two adjacent nodes, then the traffic of all types, including forwarding traffic flowing this link, can be defined as the total traffic, represented by $\widetilde{S}_{x,y}$. Let $\mathcal{Q}_{p}$ denote the set of instances deployed on node $N_{p}, p \in \mathcal{N}$, then the direct traffic between any two nodes $N_{p}$ and $N_{q}$ during a single execution of $W_{k}$ can be represented as

\vspace{-0.8em}

\begin{equation}\label{eq10}
  S_{{p},{q}}^{k} =\!\! \sum_{M_{i,a} \in \mathcal{Q}_{p}} \! \sum_{M_{j,b}\in \mathcal{Q}_{q}}
    \!\!\! \left(\widetilde{{F}}_{M_{i,a},M_{j,b}}^{k} \!+\! \widetilde{{R}}_{M_{i,a},M_{j,b}}^{k}\!\right), \forall k \in \mathcal{K}.
\end{equation}

\par Then, the direct traffic from $N_{p}$ to $N_{q}$ in unit time can be represented by

\vspace{-0.5em}

\begin{equation}\label{eq11}
   S_{p,q} = \sum\limits_{k \in \mathcal{K}} \lambda^{k} S_{{p},{q}}^{k}.
\end{equation}

\par We use $U_{p,q}$ to represent the routing path from $N_{p}$ to $N_{q}$, where $p, q \in \mathcal{N}$. Given any two adjacent nodes $N_{x}$ and $N_{y}$, if the directly connected link between them is one hop of the routing path $U_{p,q}$, then we can denote it as $\langle N_{x}, N_{y}\rangle \in U_{p,q}$. The total traffic from $N_{x}$ to its adjacent nodes $N_{y}$ can be expressed as

\vspace{-0.5em}

\begin{equation}\label{eq12}
    \widetilde{S}_{{x,y}}=
    \sum_{p = 1}^{\left|\mathcal{N}\right|} \sum_{q = 1}^{\left|\mathcal{N}\right|}
    \bigg \{
    S_{p,q},
    \ \text{if} \ \left < N_{x},N_{y} \right > \in
    {U}_{p,q} \
    \text{else} \ 0 \bigg \},
\end{equation}
where $(\ref{eq12})$ reveals that when traversing the routing path $U_{p,q}$, if the direct traffic from $N_{p}$ to $N_{q}$ is $S_{p,q}$, and $N_{x}$ to $N_{y}$ is one hop of $U_{p,q}$, then the total traffic between $N_{x}$ and $N_{y}$ is incremented by $S_{p,q}$.

\vspace{-1.3em}

\subsection{Communications Delay Analysis Model}

\par To address the issue of multiple service traffic flows sharing link bandwidth, this paper leverages queuing theory to analyze available link bandwidth. We exploit the M/M/1 queuing model to model the data packet transmission process between any two adjacent nodes $N_{x}$ and $N_{y}$. Let the packet size be $s$ \emph{kB}, the bandwidth from $N_{x}$ to $N_{y}$ be $Z_{x,y}$ \emph{Mbps}, the packet arrival rate from $N_{x}$ to $N_{y}$ be $\lambda_{x,y}$ \emph{Mbps}, and the service rate of the link be $\mu_{x,y}$ \emph{Mbps}. The average sending time for each packet between $N_{x}$ and $N_{y}$ can be expressed as

\vspace{-0.5em}

\begin{equation}\label{eq13}
   T^{pkg}_{{x,y}}= \frac{1}{\mu _{{x,y}}-\lambda _{{x,y}}}.
\end{equation}

\par The average available bandwidth of the directly connected link from $N_{x}$ to its adjacent node $N_{y}$ can be represented by

\vspace{-0.5em}

\begin{small}
\begin{equation}\label{eq14}
   \begin{aligned}
    & Z^{\prime}_{x,y} =\frac{s}{T^{pkg}_{{x,y}}}=
    s \times \bigg({\mu _{{x,y}}-\lambda _{{x,y}}}\bigg)\\
    & = s\times \bigg(\frac{Z_{x,y}}{s} - \frac{\widetilde{S}_{{x,y}}}{s}\bigg) =Z_{x,y}-\widetilde{S}_{{x,y}}.
   \end{aligned}
\end{equation}
\end{small}

\par Given any two adjacent nodes $N_{x}$ and $N_{y}$, the propagation delay between them is represented as $V_{x,y}$. Assume full-duplex communication method among nodes, the total propagation delay between any two nodes $N_{p}$ and $N_{q}$ can be expressed as follows:

\vspace{-0.5em}

\begin{equation}\label{eq15}
    \widetilde{V}_{p,q} = \sum_{\left\langle N_{x},N_{y} \right\rangle \in {U}_{p,q}} V_{x,y}.
\end{equation}

\par Let $Z_{min}^{p,q}$ denote the minimum available bandwidth among all paths from $N_{p}$ to $N_{q}$. Ignoring the forwarding time at intermediate nodes, the average communication delay $T_{M_{i,a},M_{j,b}}$ for instance $M_{i,a}$ to send requests to instance $M_{j,b}$ can be calculated as

\vspace{-0.8em}

\begin{equation}\label{eq16}
    \!\! T_{M_{i,a},M_{j,b}} =\!\!
    \begin{cases}
        \widetilde{V}_{{L_{i,a},L_{j,b}}}
        \!+\!
        \frac{D^{req}_{i,j}}
        {Z^{min}_{{L_{i,a},L_{j,b}}}}
        \!\! &, \text{ if }  L_{{i,a}} \neq L_{{j,b}},
        \\
        0 \!\! &, \text{ if } L_{{i,a}} = L_{{j,b}},
    \end{cases}
\end{equation}
where $L_{i,a}, a \in \mathcal{A}_{i}$ and $L_{j,b}, b \in \mathcal{A}_{j}$ represent the computing nodes where $M_{i,a}$ and $M_{j,b}$ are deployed, respectively. Similarly, the average communication delay for $M_{i,a}$ to send responses to $M_{j,b}$ can be given as follows:

\vspace{-0.8em}

\begin{equation}\label{eq17}
    \!\! T^{\prime}_{M_{i,a},M_{j,b}} \!=\!
    \begin{cases}
        \widetilde{V}_{{L_{i,a},L_{j,b}}}
        \!+\!
        \frac{D^{res}_{i,j}}
        {Z^{min}_{{L_{i,a},L_{j,b}}} }
          \!\! &,
        \text{ if }  L_{M_{i,a}} \neq L_{M_{j,b}}, \\
        0 \!\! &, \text{ if } L_{M_{i,a}} = L_{M_{j,b}}.
    \end{cases}
\end{equation}

\par Therefore, the average communication delay for each execution of service $W_{k}$ can be expressed as

\begin{small}
\begin{equation}\label{eq18}
    \begin{aligned}
        T_{k} = &
        \sum_{i = 0}^{\left|\mathcal{I}\right|}
        \sum_{a = 1}^{\left|\mathcal{A}_{i}\right|}
        \sum_{j = 0}^{\left|\mathcal{I}\right|}
        \sum_{b = 1}^{\left|\mathcal{A}_{j}\right|}
        \! \bigg({F}_{M_{i,a},M_{j,b}}^{k}
        T_{M_{i,a},M_{j,b}} \\
        & + {R}_{M_{i,a},M_{j,b}}^{k}
        T^{\prime}_{M_{i,a},M_{j,b}}\! \bigg)
        , \forall i \neq j, k \in \mathcal{K}.
    \end{aligned}
\end{equation}
\end{small}

\par Due to the various delay requirements of different service types, we employ the weighted sum of the average communication delays for each type of service to indicate system's overall average communication delay, which can be expressed as follows:

\vspace{-0.5em}

\begin{equation}\label{eq19}
    T = \sum_{k \in \mathcal{K}} \theta _{k} T_{k},
    \vspace{-0.8em}
\end{equation}
where $\theta_{k} \in \Theta \triangleq \left\{\theta_{1},\cdots, \theta_{K}\right\}$ denotes the weight of service $W_{k}, k \in \mathcal{K}$.

\vspace{-1em}

\section{Problem Formulation and Solutions}

\par In this section, we aim to minimize the weighted-sum of average communication delay by optimizing the microservice deployment scheme $\mathcal{L} \triangleq \left\{L_{i,a}\right\}_{i \in \mathcal{I}, a \in \mathcal{A}_{i}}$ of microservice instances, where $L_{i,a}$ indicates the computing node where microservice instance $M_{i,a}$ is deployed. Based on the discussion in Sec. II A$-$D, we can formulate the minimization problem as follows:

\vspace{-1.3em}

\begin{subequations}\label{P1}
    \begin{align}
        \mathcal{P}1: & \min_{\left\{L_{i,a}\right\}_{i \in \mathcal{I}, a \in \mathcal{A}_{i}}} \ \  T = \sum_{k\in \mathcal{K}} \theta _{k} T_{k}, \label{e20a}                                       \\
        \text{s.t.}\
                               & \sum _{M_{i,a} \in \mathcal{M}_{p}}C_{i} \leqslant C_{p}, \ \forall N_{p} \in \mathcal{N} \label{e20b}, \\
                               & \sum _{M_{i,a} \in \mathcal{M}_{p}}B_{i} \leqslant B_{p},\ \forall N_{p} \in \mathcal{N} \label{e20c},   \\
                               & \widetilde{S} _{{p},{q}} \leqslant Z_{p,q}, \ \forall q \in \mathcal{N}_{p}, \ \forall p \in \mathcal{N} \label{e20d},
    \end{align}
\end{subequations}
where $\mathcal{M}_{p}$ and $\mathcal{N}_{p}$ represents the sets of all instances of microservices deployed and the nodes adjacent to node $N_{p}, p \in \mathcal{N}$, respectively. $C_{i}$ and $B_{i}$ indicate the computational and memory resources required to deploy the instance of microservice $M_{i}$ on node $N_{p}$, respectively. Constraints (\ref{e20b}) and (\ref{e20c}) guarantee that the total computational and memory resources required by the microservice instances on each node do not exceed the node's CPU count and memory capacity, respectively. Constraint (\ref{e20d}) ensures that the traffic on each link does not exceed its bandwidth limit.

\vspace{-1em}

\subsection{Proposed Solution}
\par Due to the numerous microservices and nodes, it is extremely challenging to efficiently tackle the optimal solution of the original problem $\mathcal{P}1$ leveraging traditional convex optimization methods. To this end, we first introduce $0$-$1$ variables to transform $\mathcal{P}1$ into a nonlinear $0$-$1$ programming problem.

\subsubsection{Problem Transformation} Generally, to guarantee the availability of services in edge networks, instances of each microservice should be deployed across multiple nodes. In this case, we introduce binary variables $G_{p,i} \in \mathcal{G}$ to characterize whether the instance of microservice $M_{i}, i \in \mathcal{I}$ is deployed on node $N_{p}$. Moreover, we use $M_{p,i}$ to denote the instance of microservice $M_{i}$ located on node $N_{p}$, thereby uniquely identifying whether the instance of $M_{i}$ is on $N_{p} \in \mathcal{N}$ or not. Then, (\ref{eq10}) can be reformulated as follows:

\vspace{-1em}

\begin{equation}\label{e21}
    S_{{p},{q}}^{k}=\sum_{M_{i} \in \mathcal{M}} \sum_{M_{j}\in \mathcal{M}}
    G_{p,i}G_{q,j}
    (\widetilde{{F}}_{M_{p,i},M_{q,j}}^{k} + \widetilde{{R}}_{M_{p,i},M_{q,j}}^{k})
    .
\end{equation}

\par For the topology relationships, we introduce binary variable $I_{p,q}^{x,y} \in I_{p,q}$ to indicate whether the path from node $N_{p}$ to node $N_{q}$ includes the hop between adjacent nodes $N_{x}$ and $N_{y}$. Then, (\ref{eq12}) can be reformulated as follows:

\vspace{-0.8em}

\begin{small}
\begin{equation}\label{e22}
    \widetilde{S}_{{x,y}}=
    \sum_{p=1}^{\left|\mathcal{N}\right|} \sum_{q=1}^{\left|\mathcal{N}\right|}
    I^{p,q}_{x,y} S_{{p}{q}}.
\end{equation}
\end{small}

\par Similarly, (\ref{eq15}) can also be rewritten as follows:

\vspace{-0.5em}

\begin{equation}\label{e23}
    \widetilde{V}_{p,q}=\sum_{I_{p,q}^{x,y} \in I_{p,q}} I_{p,q}^{x,y} V_{x,y}.
\end{equation}

\par Furthermore, (\ref{eq16}) and (\ref{eq17}) can be reformulated as follows:

\vspace{-0.5em}

\begin{small}
\begin{equation}\label{e24}
    T_{M_{p,i},M_{q,j}}=
    \begin{cases}
        \widetilde{V}_{p,q} + \frac{D^{req}_{i,j}}{Z^{min}_{p,q}} &, \text{if}\ p \neq q \wedge i \neq j,\\
        0 &, \text{else},
    \end{cases}
\end{equation}
\end{small}

\vspace{-0.5em}

\begin{small}
\begin{equation}\label{eq25}
    T^{\prime}_{M_{p,i},M_{q,j}}=
    \begin{cases}
        \widetilde{V}_{p,q}+\frac{D^{res}_{i,j}}{Z^{min}_{p,q}}&, \text{if}\ p \neq q \wedge i\neq j,\\
        0 &, \text{ else }.
    \end{cases}
\end{equation}
\end{small}

\par Finally, (\ref{eq18}) can be can be transformed as follows:

\vspace{-0.5em}

\begin{equation}\label{eq26}
    \begin{aligned}
        T_{k}= & \sum_{i=0}^{\left|\mathcal{I}\right|}
        \sum_{p=1}^{\left|\mathcal{N}\right|}
        \sum_{j=0}^{\left|\mathcal{I}\right|}
        \sum_{q=1}^{\left|\mathcal{N}\right|}
        G_{p,i}G_{q,j}
        \bigg ({F}_{M_{p,i},M_{q,j}}^{k}
        T_{M_{p,i},M_{q,j}}\\
        & + {R}_{M_{p,i},M_{q,j}}^{k}
        T'_{M_{p,i},M_{q,j}}\bigg)
        , \ i \neq j.
    \end{aligned}
\end{equation}

\subsubsection{Problem Reformulation} As a result, the original problem $\mathcal{P}1$ can be equivalently transformed into the following form:

\vspace{-1.5em}

\begin{subequations}\label{eq27}
    \begin{align}
        \mathcal{P}2: & \min_{\mathcal{G} : \left\{G_{p,i}\right\}_{i \in \mathcal{I}, p \in \mathcal{N}}} \ \  T = \sum_{W_{k} \in \mathcal{W}} \theta_{k} T_{k}, \label{eq27a}                               \\
        \text{s.t.}\
                      & \sum_{p = 1}^{\left|\mathcal{N}\right|} G_{p,i} \leq \left |\mathcal{A}_{i}\right |, \  \forall i \in \mathcal{I}, \label{eq27b} \\
                      & \sum_{i=1}^{\left|\mathcal{I}\right|} G_{p,i} \leq 1, \  \forall N_{p} \in \mathcal{N}, \label{eq27c}                          \\
                      & \sum_{i=1}^{\left|\mathcal{I}\right|} G_{p,i}C_i\leqslant C_{p}, \ \forall p \in \mathcal{N},  \label{eq27d}               \\ 
                      & \sum_{i=1}^{\left|\mathcal{I}\right|}G_{p,i}B_i\leqslant B_{p}, \ \forall p \in \mathcal{N},  \label{eq27e}               \\
                      & \widetilde{\Gamma } _{{p},{q}} \leqslant Z_{p,q}, \ \forall q \in \mathcal{N}_{p}, \ \forall p \in \mathcal{N}, \label{eq27f}
    \end{align}
\end{subequations}
where $(\ref{eq27b})$ and $(\ref{eq27c})$ specify the constraints on the number of microservice instances. In particular, $(\ref{eq27b})$ ensures that the total instances of microservice $M_{i}$ across do not exceed  $\left|\mathcal{A}_{i}\right|$. $(\ref{eq27c})$ prevents multiple instances of $M_{i}$ from being deployed on the same node. Constraints $(\ref{eq27d})$ and $(\ref{eq27e})$ ensure that the computational and memory resources required by the deployed microservices $M_{i}$ cannot exceed the total computational and memory resources available on the node. Constraint $(\ref{eq27f})$ enforces the bandwidth constraint between node $N_{p}$ and its adjacent nodes $N_{q}$.

\vspace{-1em}

\subsection{Topology-aware and Individual-adaptive Microservice Deployment Scheme}
\par Given a large number of microservices and nodes, effectively tackling the nonlinear $0$-$1$ programming problem $\mathcal{P}2$ optimally within a constrained timeframe is challenging. While the genetic algorithm (GA) offers a robust heuristic approach for combinatorial problems, its inherent limitations of slow convergence and suboptimal performance in specific scenarios necessitate refinement. To address these shortcomings, we propose a novel topology-aware and individual-adaptive microservice deployment (TAIA-MD) scheme. This scheme leverages the perceived network topology to design the microservice deployment scheme. Moreover, it incorporates an individual-adaptive mechanism that enhances individual adaptability by strategically initializing a select group of ``super individuals'' within the GA population, thereby accelerating convergence and avoiding local optima. As described in \textbf{Algorithm 1}, the first part of our proposed TAIA-MD scheme encompasses four aspects, as follows:

\begin{itemize}
  \item \textbf{Chromosome Encoding:} We employ binary encoding, where each chromosome consists of $\left|\mathcal{I}\right|$ genes, representing the deployment of $\left|\mathcal{I}\right|$ microservice types. The set of computing nodes available for deployment is $\left|\mathcal{N}_{cmp}\right|$. Each gene is a binary string of length $\left|\mathcal{N}_{cmp}\right|$, with a value of `$0$' or `$1$' indicating whether the microservice is deployed on the corresponding computing node or not. In addition, the number of `$1$'s on each gene is consistent with the number of instances of the corresponding microservice type.

  \item \textbf{Fitness Function:} The fitness function measures the quality of an individual in solving $\mathcal{P}2$. To minimize the average communication delay $T$, the fitness function is defined as a sufficiently large number minus $T$. The calculation of $T$ is detailed in \textbf{Algorithm 2}.

  \item \textbf{Genetic Operators:} Genetic operators generate new individuals through selection, crossover, and mutation. For the selection operation, we implement the tournament method as our selection scheme. Specifically, two groups of individuals are randomly sampled from the population, with each containing $Y_{m}$ individuals. The fittest individual from each group is then selected as the parent. Following the selection operation, crossover operations are performed on these selected parents. Each gene from the parent has a probability of $Y_{p}$ to exchange two genes if the chromosomes of both individuals are feasible after exchanging. Subsequently, the offspring undergo mutation with a probability of $Y_{q}$, where bits are randomly flipped to maintain chromosome feasibility. The process of selection, crossover, and mutation repeats until the offspring population reaches the desired population size $Y_{n}$.

  \item \textbf{Termination Conditions:} The algorithm has a maximum iteration count of $Y_{k}$ to ensure timely problem-solving. Termination transpires upon reaching this threshold or when individual fitness for $Y_{i}$ consecutive iterations has no significant change, conserving computational resources.
\end{itemize}

\setlength{\textfloatsep}{1pt}

\begin{algorithm}
    \caption{Topology-aware genetic algorithm}
    \setstretch{0.4}
    \label{alg:TGA}
    \KwIn{Service set $\mathcal{W}$; Microservice set $\mathcal{M}$; Node set $\mathcal{N}$; Population size $Y_n$; Number of individuals per group $Y_m$; Gene crossover probability $Y_p$; genetic mutation probability $Y_q$; Maximum number of iterations $Y_k$; Current number of iterations $Y_i$}
    \KwOut{Optimal deployment scheme $\mathcal{G}^{best}$}
    Randomly initialize each individual $I$ in the population $\mathbb{P}$;\\
    \For{$\text{i}\gets\text{1}~\textbf{to}~Y_k$}{
        \ForEach{Individual $I$ in $\mathbb{P}$}{
            For the deployment scheme $\mathcal{G}$ of individual $I$, compute its time delay $T$;\\
            Updating the fitness of individual $I$;\\
        }
        Update the optimal scheme $\mathcal{G}^{best}$;\\
        \If{The un-updated round of $\mathcal{G}^{best}$ $>$ $Y_i$}{
            End the loop;\\
        }
        Create a collection of child individuals $\mathbb{I}'$;\\

        \While{$\left|\mathbb{I}'\right|<Y_n$}{
            Randomly select two groups of individuals, each with the number of $Y_m$;\\
            Select the top two individuals with the highest fitness from each group, $I_A$ and $I_B$, as parents;\\
            $I_A$ and $I_B$ undergo uniform crossover with probability $Y_{p}$, producing offspring $I'_A$ and $I'_B$;\\
            Each gene on $I'_A$ and $I'_B$ undergoes mutation with probability $Y_{q}$;\\
            Add $I'_A$ and $I'_B$ into $\mathbb{I}'$;\\
        }
        Replace the population of individuals with $\mathbb{I}'$;\\
    }
\end{algorithm}

\setlength{\textfloatsep}{1pt}

\begin{algorithm}
    \caption{Microservice Deployment Scheme Evaluation Algorithm}
    \setstretch{0.35}
    \label{alg:Scheme_Evaluation}
    \KwIn{Microservice deployment scheme $\mathcal{G}$; Service set $\mathcal{W}$; Microservice set $\mathcal{M}$; Node set $\mathcal{N}$; Service rating weight set $\Theta$}
    \KwOut{System average communication delay $T$}
     Compute the propagation delay $\widetilde{V}_{p,q}$ between any nodes;\\
    \ForEach{Service $W_k$ in $\mathcal{W}$}{
     Compute $F_{M_{i,a},M_{j,b}}^{k}$ and $R_{M_{i,a},M_{j,b}}^{k}$ between any two instances during one execution of $W_k$;\\
    }
    Compute the direct traffic $S_{p,q}$ between any two nodes;\\
    Compute the total traffic $\widetilde{S}_{x,y}$ between any adjacent nodes;\\
    Compute the actual available bandwidth $Z'_{x,y}$ between any adjacent nodes;
    Compute the minimum available bandwidth $Z^{min}_{p,q}$ between any nodes;\\
    Initialize the system average communication delay $T$;\\

    \ForEach{Service $W_k$ in $\mathcal{W}$}{
    Initialize the average communication delay $T_k$ of service $W_k$;\\
    \ForEach{Invoke $M_{i}, M_{j}$ during the execution of service $W_{k}$}{

    Denote the node sets of deploying $M_i$ and $M_j$ as $\mathcal{N}^i$ and $\mathcal{N}^j$, respectively;\\

    \ForEach{$N_p\in \mathcal{N}^i$}{
    \ForEach{$N_q\in \mathcal{N}^j$}{

    Compute $T_{M_{i,a},M_{j,b}}$ and $T^{\prime}_{M_{i,a},M_{j,b}}$ between the corresponding instances on two nodes;\\
    $T_k$ += $F _{M_{i,a},M_{j,b}}^{k}T_{M_{i,a},M_{j,b}} + R _{M_{i,a},M_{j,b}}^{k}T^{\prime}_{M_{i,a},M_{j,b}}$;\\
    }

    }
    }
    T += $\theta _{k} T_k$;\\
    }
\end{algorithm}


\par Moreover, we implement an individual-adaptive mechanism. This mechanism generates a select few ``super individuals'' during population initialization and endows them with adaptive capabilities to accelerate algorithm convergence. \textbf{Algorithm 3} introduces a low-complexity algorithm named greedy-based time optimization algorithm to fine-tune deployment strategies for these super individuals. By altering the traversal order of services in line 3 of \textbf{Algorithm 3}, we can generate diverse deployment schemes, ensuring the genetic diversity of super individuals. To prevent the algorithm from getting trapped in local optima, we strictly limit the number of super individuals. Coupled with the tournament selection scheme, non-super individuals still have a significant chance of being selected as parents.

\vspace{-1em}

\subsection{Computational Complexity Analysis}
\par We first denote the maximum number of instances for each type of microservice as $X$. Then the complexity for randomly generating the deployment scheme in the initialization phase can be calculated $\mathcal{O}\left(X\left|\mathcal{M}\right|\right)$. In \textbf{Algorithm 3}, the complexity of deploying each microservice instance based on the invocation relationship is $\mathcal{O}\left(X\left|\mathcal{M}\right|\left|\mathcal{N}\right|\right)$. As illustrated in \textbf{Algorithm 1}, each iteration involves generating a new population and evaluating its fitness. As a result, the complexities for selection, crossover, and mutation operations are $\mathcal{O}\left(Y_{m}\right)$, $\mathcal{O}\left(\left|\mathcal{M}\right|\left|\mathcal{N}\right|\right)$, and $\mathcal{O}\left(1\right)$, respectively, resulting in the complexity for producing a new individual of $\mathcal{O}\left(Y_{m} + \left|\mathcal{M}\right|\left|\mathcal{N}\right|\right)$. The fitness evaluation in \textbf{Algorithm 2} considers node-to-node forwarding paths and microservice instance invocation relationships, leading to a complexity of $\mathcal{O}\left(X^{2}\left|\mathcal{M}\right|^{2}\right)$. \textbf{Algorithm 2} may require up to $Y_{k}$ iterations, each producing $Y_{n}$ individuals and evaluating their fitness. Compared to the iterative process, the initialization phase's complexity is negligible. As a result, the proposed TASA-MD scheme has an overall computational complexity of $\mathcal{O}\left(Y_{k}Y_{n}\big(Y_{m} + X^{2}\left|\mathcal{M}\right|^{2}\big)\right)$, significantly improving solving efficiency compared to the exponential time complexity of the enumeration algorithm $\mathcal{O}\left(\big(X\left|\mathcal{M}\right|\big)^{\left|\mathcal{N}\right|}\right)$.

\setlength{\textfloatsep}{1pt}

\begin{algorithm}
    \caption{Greedy-based Microservice Deployment Algorithm}
    \setstretch{0.35}
    \label{alg:Greddy}
    \KwIn{Service set $\mathcal{W}$; Microservice set $\mathcal{M}$; Node set $\mathcal{N}$.}
    \KwOut{Microservice deployment scheme $\mathcal{G}$.}
    Initialize microservice deployment scheme $\mathcal{G}$;\\
    Compute the communication delay between any two nodes $N_p$ and $N_q$ in $\mathcal{N}$;
    \ForEach{$W_k$ in $\mathcal{W}$}{
    \If{$M_i$ is not deployed}{
        Compute the average delay from each computing node to all access nodes;\\
        Deploy instances of $M_i$ to $\left|\mathcal{A}_{i}\right|$ nodes with the lowest average delay that meet the resource requirements of $M_i$;\\
        Update the remaining resources of the corresponding computing nodes;\\
        }
        \ForEach{$M_i$ required by $W_k$}{
        \If{$M_i$ is not deployed}{
        Denote the set of associated nodes deployed with $M_i$ frontend $M_j$ as $\mathcal{N}'$;\\
        Compute the average delay from each computing node to the nodes in $\mathcal{N}'$;\\
        Deploy instances of $M_i$ to the $\left|\mathcal{A}_{i}\right|$ nodes with the lowest average delay that meet resource requirements of $M_i$;\\
        Update the remaining resources of the corresponding computing nodes;\\
        }
        }
    }
\end{algorithm}

\vspace{-0.5em}

\section{Performance Evaluation}

\vspace{-0.5em}

\par To comprehensively validate the superior performance of the proposed TAIA-MD scheme, we have conducted extensive simulations focusing on bandwidth resources, service arrival rates, computing resources, and network topology. In addition, to further demonstrate the effectiveness and practicality of the TAIA-MD scheme, we have also carried out the actual deployment on the previously developed microservice physical verification platform called DMSA \cite{yuangchen_DMSA}. These evaluations collectively confirm the robustness and adaptability of the TAIA-MD scheme in both simulated and real-world environments Notably, due to access restrictions on deployment permissions on commercial platforms such as Amazon and Netflix, our current performance evaluation, like most studies, is independent of actual and commercial parameters..

\vspace{-1em}

\subsection{Simulation Parameter Settings}

\vspace{-0.4em}

\par In our simulations, we consider there are $50$ types of microservices, each with input and output data sizes ranging from $10$ KB to $100$ KB, requiring $0.1$ to $0.3$ CPU units and $100$ MB to $300$ MB of memory per instance, typically deployed in $3$ to $5$ instances. There are $10$ service types with $5 \! \sim \! 8$ microservice invocations and an arrival rate $\lambda^{k}$ of $30 \! \sim \! 50$. The edge network topology comprises $50$ nodes, including $35$ computing nodes, $10$ routing nodes, and $5$ user-access nodes. Each link has a propagation delay of $10 \mu s$ and a bandwidth of $100$ Mbps. Microservice instances are deployed on computing nodes, user requests arrive via user-access nodes, and routing nodes handle data forwarding. Each computing node has $8$ CPU cores and $8000$ MB memory. Notably, the simulation parameter settings are randomly generated within a reasonable range with reference to works \cite{9419855,9740415,9154603} as well as the edge environment characteristics.

\par Furthermore, since different network topologies can significantly impact the performance of edge networks, for example, links used by multiple routing paths may lead to bandwidth shortages, in order to quantify this, we introduce the concept of link forwarding load (LFL) to indicate how often a link serves as the forwarding path in the network topology.

\begin{myDef}\label{def1}
  Link forwarding load (LFL): For any given $N_{x}$ and its adjacent node $N_{y}$, the link forwarding load between them can be represented as $O_{x,y} = \sum_{p = 1}^{\left|\mathcal{N}\right|}\sum_{q = 1}^{\left|\mathcal{N}\right|} I_{p,q}^{x,y}$. Let $\mathbb{E}$ denote the set of edges in the network topology, then the average link forwarding load of the network topology can be represented as
  \begin{equation}\label{eq28}
     O_{avg} = \frac{\sum_{x=1}^{\left|\mathcal{N}\right|} \sum_{y=1}^{\left|\mathcal{N}\right|} O_{x,y}}{\left|\mathbb{E}\right|}.
  \end{equation}
\end{myDef}

\begin{figure}[h]
\vspace{-0.5em}
\centering
\includegraphics[scale=0.3]{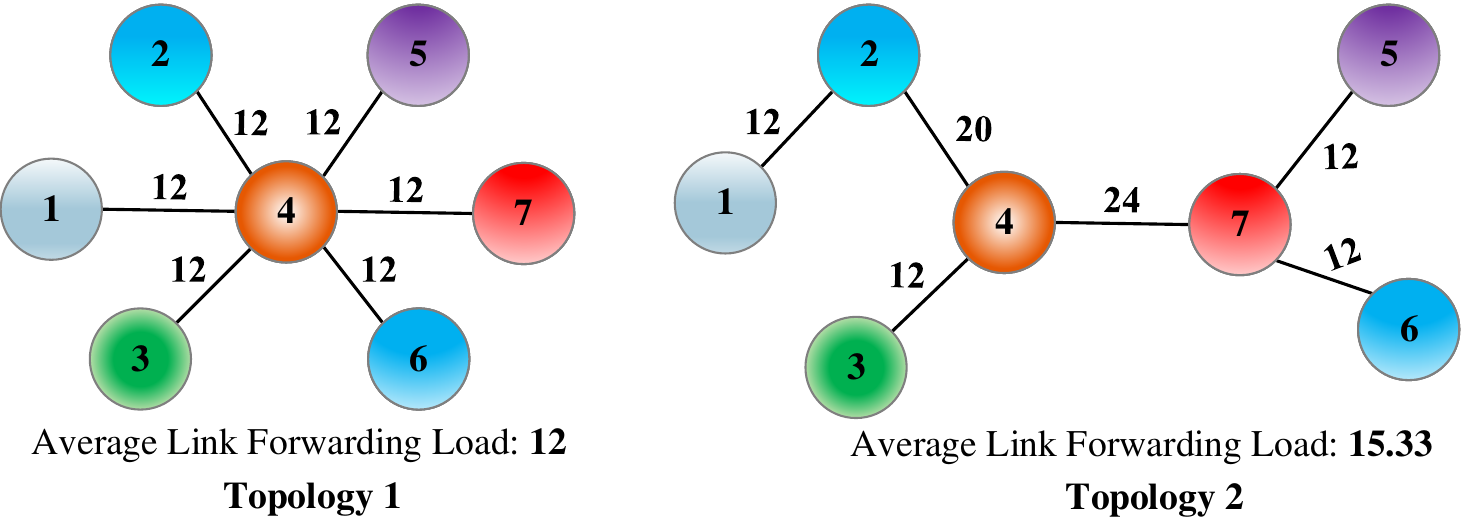}
\vspace{-0.6em}
\caption{\small Example of LFLs for two different topologies.}
\label{fig6}
\vspace{-0.5em}
\end{figure}

\par
\begin{figure}[t]
\vspace{-0.5em}
\centering
\includegraphics[scale=0.35]{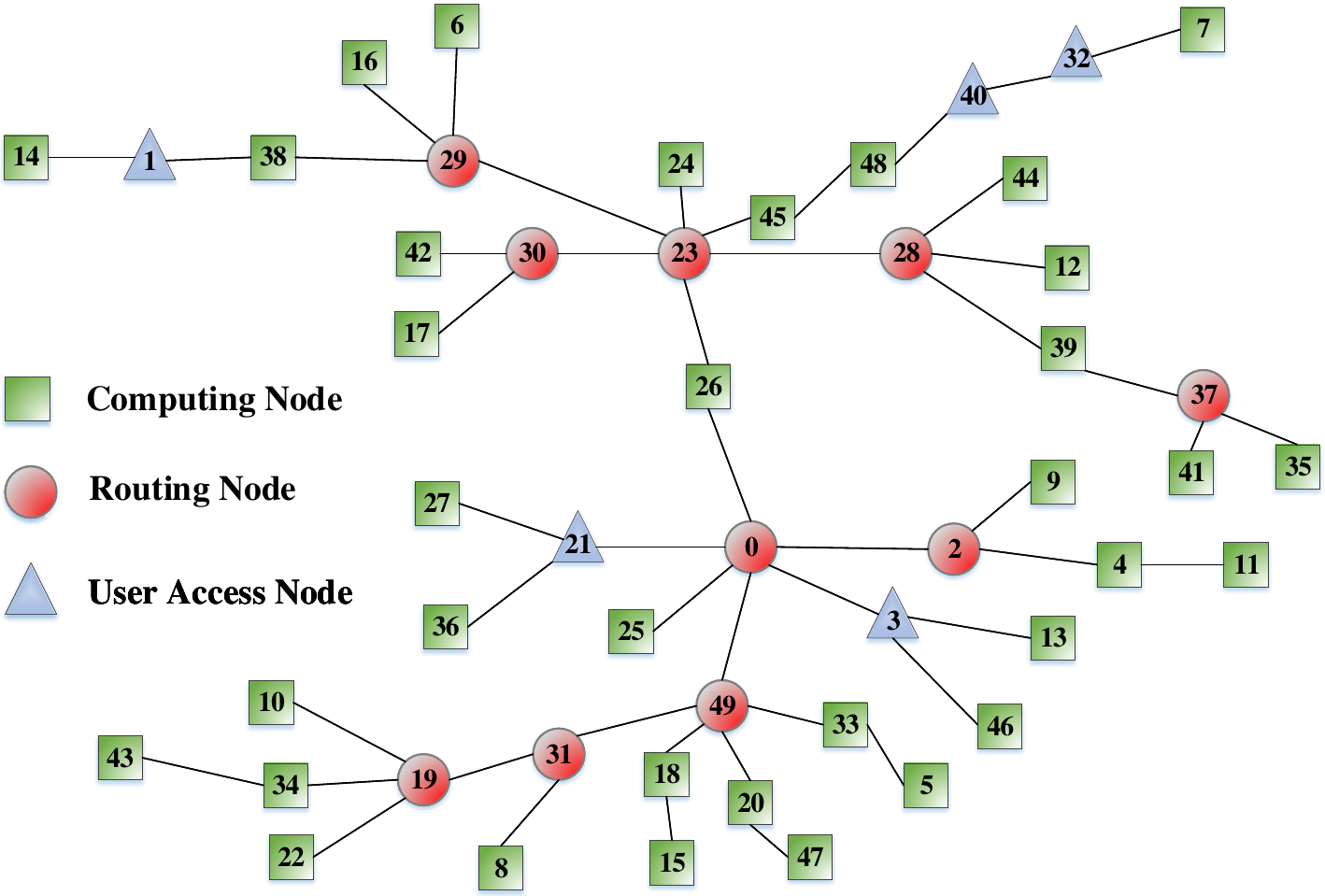}
\vspace{-0.6em}
\caption{\small Default topology used in the simulations.}
\label{fig7}
\end{figure}

 Fig. \ref{fig6} depicts the LFLs across two different topologies, where edge weights signify LFLs calculated per \textbf{Definition \ref{def1}}. In Topology 1, all adjacent nodes have LFLs of $12$, resulting in an average LFL of $12$. In Topology 2, links such as $E_{2,4}$ and $E_{4,7}$ bear higher routing tasks, leading to higher LFLs. Despite both topologies having identical node and edge counts, Topology 2 exhibits a higher average LFL, making it more prone to bandwidth bottlenecks. In our simulation's default topology, the connections between nodes in the edge network are based on the concept of the defined LFL. By adjusting different LFLs, we can simulate various network topologies. The default topology used in our simulations is shown in Fig. \ref{fig7}.

\vspace{-0.6em}

\subsection{The Setup of Physical Verification Platform DMSA}

\par
\begin{figure}[h]
    \centering
    \begin{minipage}[b]{0.48\textwidth}
        \centering
        \begin{subfigure}{0.48\columnwidth}
            \centering
            \includegraphics[width=\linewidth]{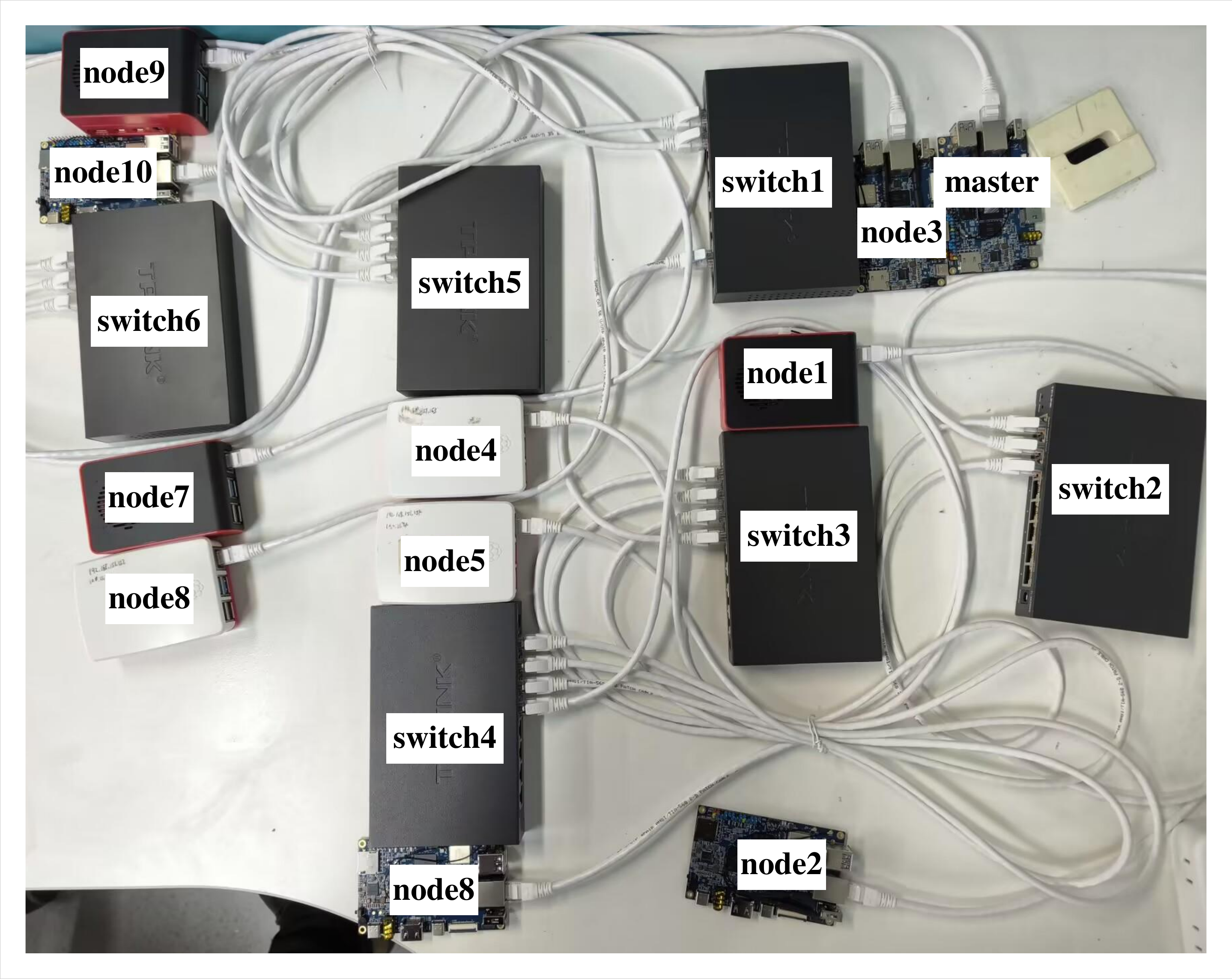}
            \caption{Practical platform of DMSA}
        \end{subfigure}
        \hfill
        \begin{subfigure}{0.48\columnwidth}
            \centering
            \includegraphics[width=\linewidth]{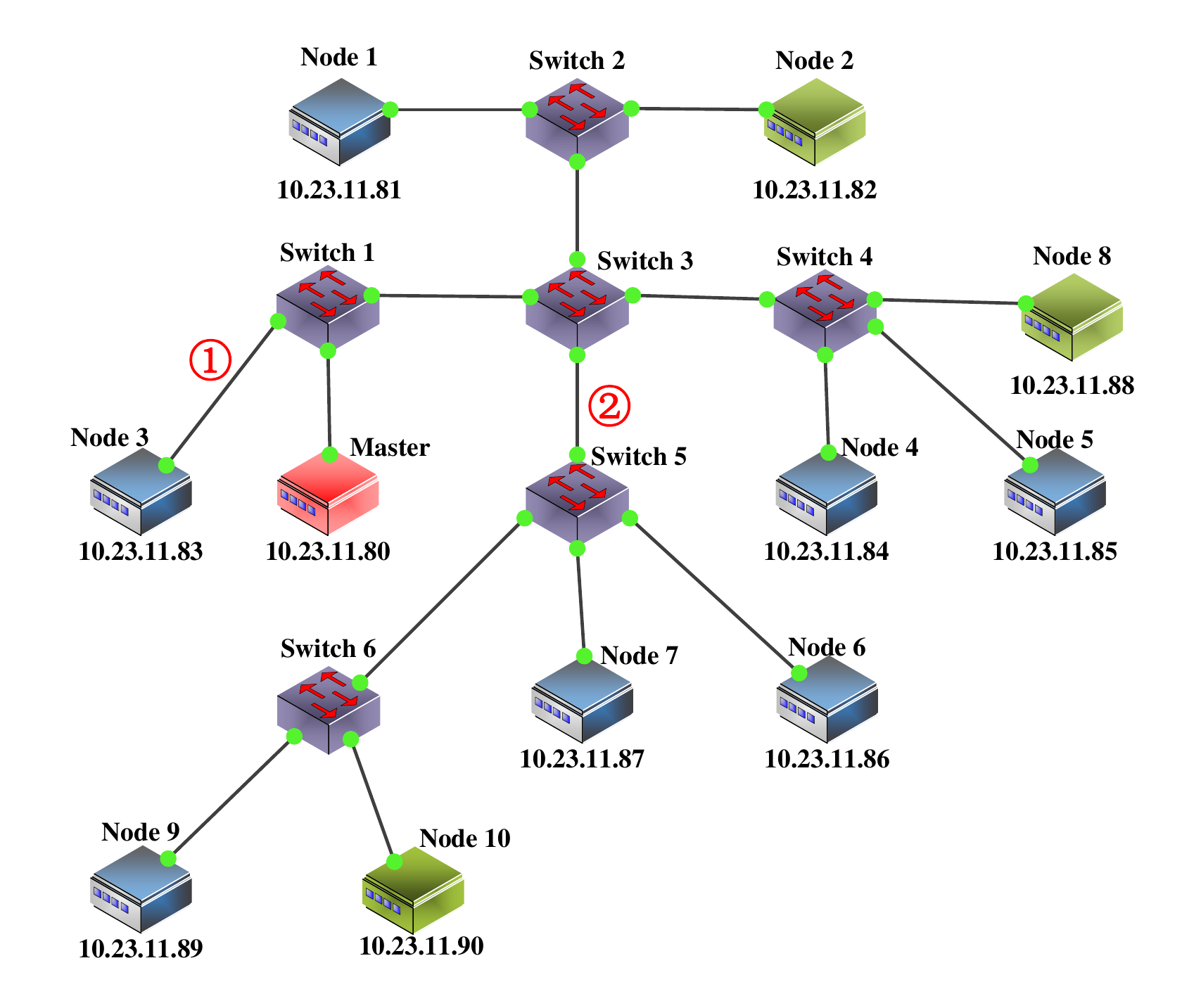}
            \caption{Logical topology of DMSA}
        \end{subfigure}
        \vspace{-0.3em}
        \caption{\small The real-world physical verification platform of DMSA, (a) practical platform of DMSA, and (b) logical topology of DMSA.}
        \vspace{0.3em}
        \label{fig_dmsa}
    \end{minipage}
    \vspace{-0.8em}
\end{figure}

 DMSA is a decentralized MSA platform, and the biggest difference from traditional centralized MSAs is that it sinks the scheduling function from the control plane to edge nodes \cite{yuangchen_DMSA}. In particular, DMSA has redesigned three core modules of microservice discovery, monitoring, and scheduling, which achieve precise awareness of instance deployments, low monitoring overhead and measurement errors, and accurate dynamic scheduling, respectively. However, despite having a comprehensive scheduling mechanism, DMSA has not yet integrated microservice deployment optimization \cite{yuangchen_DMSA}. The picture of the real-world physical verification platform DMSA we constructed is shown in Fig. \ref{fig_dmsa} (a).

\par In this paper, we have practically deployed the TAIA-MD scheme on the DMSA platform and verified its effectiveness and practicality. In particular, we exploit 6 \texttt{Raspberry Pi 4B}s, 5 \texttt{Orange Pi 5B}s, and 6 \texttt{Gigabit Ethernet Switch}es to construct this edge network topology. This network topology includes 17 nodes, namely 6 \texttt{Gigabit Ethernet} switches as communication nodes, 3 \texttt{Raspberry Pi 4B} as user-access nodes, and 9 computing nodes (4 \texttt{Raspberry Pi 4B} and 5 \texttt{Orange Pi 5B}). We have designed three typical services: video services, download services, and graphic-text services. These services are implemented by 10 microservice instances \footnote{Notably, three user-access nodes generate requests for three types of services at a specific arrival rate with loads evenly distributed.}. Video segments range from $1$ to 3 MB with a maximum wait time of $10$ seconds. Graphic-text pages are $0.5$ to $1$ MB with the same wait time, while download files are $10$ to $20$ MB with a $100$-second wait time. If it times out, the service execution will be marked as failed. Each test lasts for $40$ minutes. To comprehensively evaluate the network performance of the TAIA-MD scheme deployed on the DMSA platform, we design two network emergencies, as shown in Fig. \ref{fig_dmsa} (b). {\Large \ding{192}} Link 1 between Switch 1 and Node 3 is disconnected at the 10th minute and restored at the 15th minute to simulate the computing node suddenly goes offline. {\Large \ding{193}} Link 3 between Switch 3 and Switch 5 is restricted to 100 Mbps at the 30th minute and restored back to $1$ Gbps at the 35th minute to portray the network fluctuations. As shown in Table \ref{tab1}, we also test the performance of TAIA-MD under high, medium, and low load conditions on the DMSA platform.

\begin{table}[h]
\captionsetup{font=small}
\setlength{\belowcaptionskip}{2bp}
\centering
\caption{\small Arrival Rates of Different Services under Various Loads.}
\label{tab1}
\scriptsize
\renewcommand{\arraystretch}{0.85}
\setlength{\tabcolsep}{4pt}
\begin{tabular}{lccc}
\toprule
Load Condition & Video Service & Download Service & Graphic-text Service \\
\midrule
High Load & 5 & 1 & 10 \\
Medium Load & 3 & 0.6 & 6 \\
Low Load & 1.5 & 0.3 & 3 \\
\bottomrule
\end{tabular}
\end{table}

\subsection{Comparison Schemes in Numerous Simulations}

\par To comprehensively validate the effectiveness of the proposed TAIA-MD scheme, we compare it with four different baseline schemes in simulations, as follows:
\begin{itemize}
  \item Random Deployment Scheme \textbf{(Random)} \cite{10379832}: Microservice instances are deployed randomly on selected nodes that meet resource requirements, generating the deployment scheme.

  \item Greedy-based Deployment Scheme \textbf{(Greedy)} \cite{9740415}: This scheme considers the complex dependencies among microservices, which are modeled as the chain structure but ignoring the network topology. It only considers the previous and next microservices in the invocation chain when deploying each instance, selecting the best-performing node after evaluating all possible nodes.

  \item GA-Based Deployment Scheme \textbf{(GA)} \cite{9162056}: This scheme takes into account the dependencies between microservices and the bandwidth capacity of edge nodes but overlooks the network topology. Since it also uses the GA, comparing it with our proposed TAIA-MD scheme highlights the performance gains from incorporating topology awareness.

  \item Non-Individual-Adaptive TAIA-MD Scheme \textbf{(w/o, IA) TAIA-MD}: This variant of the proposed TAIA-MD scheme omits the individual-adaptation mechanism to validate the effectiveness individual-adaptation mechanism in accelerating convergence and avoiding local optima.
\end{itemize}

\vspace{-1.5em}

\subsection{Numerous Simulation Results}
\par
\begin{figure}[h]
\vspace{-0.5em}
\centering
\includegraphics[scale=0.35]{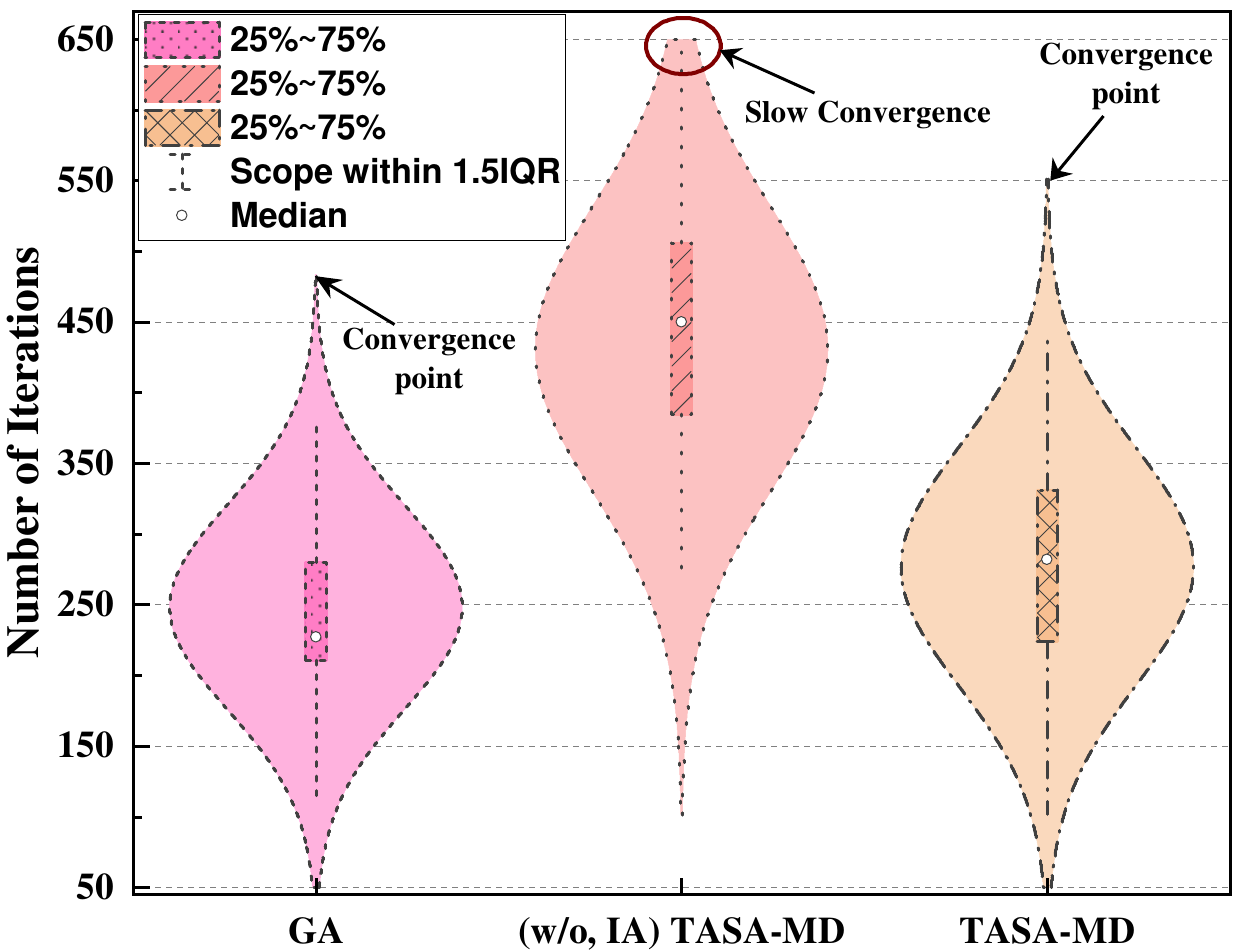}
\vspace{-0.6em}
\caption{\small Comparison of convergence performance among the proposed TAIA-MD scheme and baseline schemes.}
\label{fig9}
\end{figure}

 As illustrated in Fig. \ref{fig9}, we analyze the convergence performance of the traditional GA, (w/o, IA) TAIA-MD, and TAIA-MD schemes. Traditional GA converges quickly with fewer iterations. This primarily stems from its oversight of the edge network topology and individual adaptability, considering fewer influencing factors, thus necessitating fewer convergence iterations. Conversely, (w/o IA) TAIA-MD, which takes into account network topology and bandwidth contention, is more significantly impacted by deployment scheme variations, necessitating a longer optimization period. Addressing these issues, our proposed TAIA-MD scheme generates relatively superior individuals and gene fragments during genetic initialization, facilitating rapid convergence. Numerical results underscore that compared to (w/o, IA) TAIA-MD scheme, the proposed TAIA-MD reduces the iteration rounds by approximately $50\%$, typically completing within $300$ iterations. Furthermore, the TAIA-MD scheme tackles the tardy convergence issue of (w/o, IA) TAIA-MD, significantly enhancing algorithm efficiency. Noteworthy is that although traditional GA exhibits slightly faster convergence, its deployment schemes are far inferior compared to (w/o, IA) TAIA-MD and TAIA-MD, as will be detailed later.

\par
\begin{figure}[h]
\vspace{-0.5em}
\centering
\includegraphics[scale=0.35]{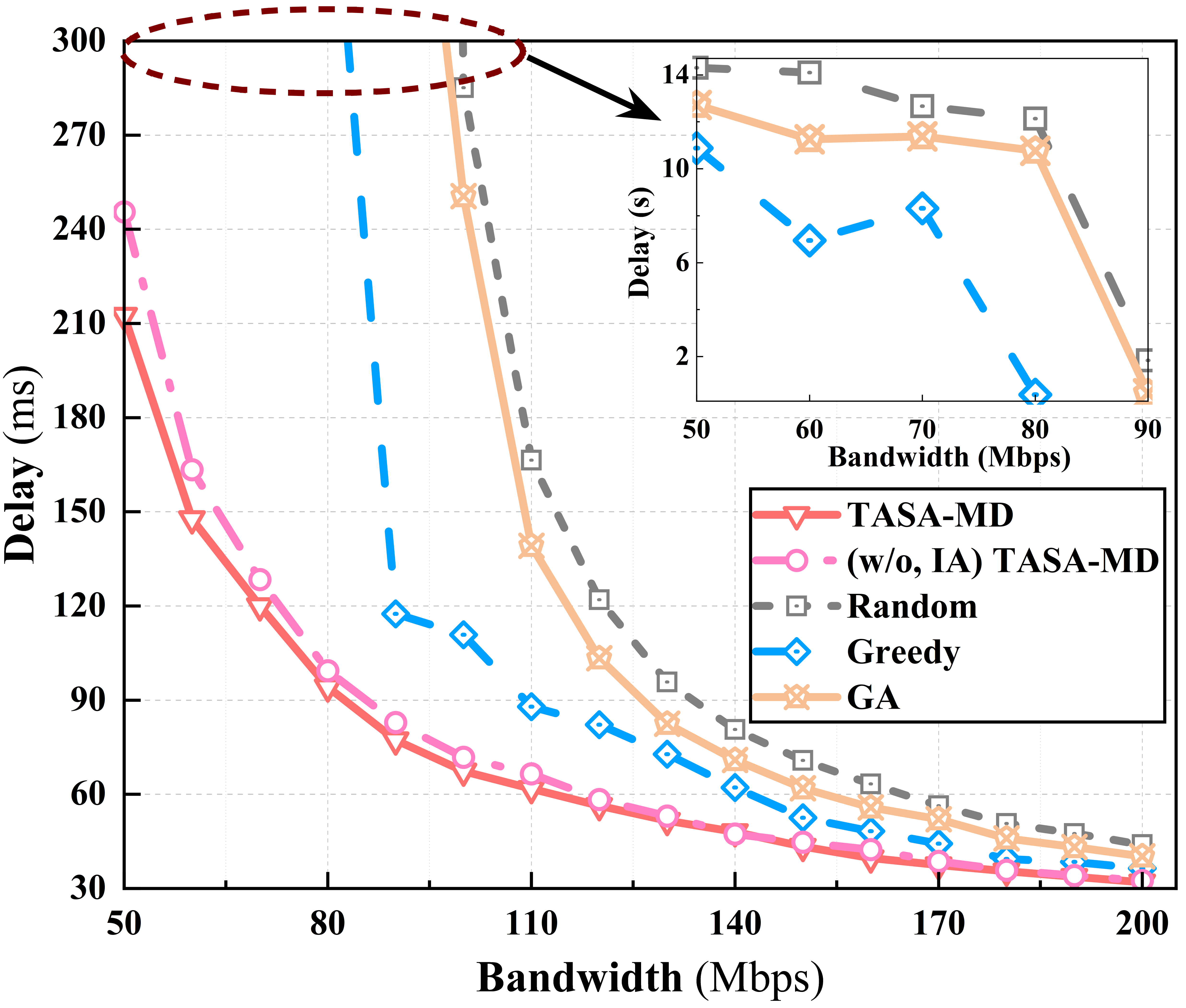}
\vspace{-0.6em}
\caption{\small Performance comparison between TAIA-MD and baseline schemes under different bandwidths.}
\label{fig10}
\end{figure}

 As depicted in Fig. \ref{fig10}, we compare the performance of TAIA-MD against the baseline schemes across different link bandwidths. In the scenarios with lower bandwidth, due to certain link traffic exceeding available bandwidth, the Random, Greedy, and GA schemes experience severe congestion and significant delays. In contrast, (w/o, IA) TAIA-MD and TAIA-MD can effectively reduce the delay by over $90\%$ and competently prevent congestion through optimized microservices deployment. As bandwidth scales up to $100$ Mbps, the congestion and delay issues for Random, Greedy, and GA schemes can be significantly alleviated. Nonetheless, (w/o, IA) TAIA-MD and TAIA-MD schemes still demonstrate remarkable performance advantages. This primarily arises from their consideration of the impact of microservice deployment on available bandwidth, incorporating network topology and individual adaptation. Compared to Greedy and GA, (w/o, IA) TAIA-MD reduces delay by $73.1\%$ and $39.3\%$, respectively. As the link bandwidth further increases to above $150$ Mbps, due to the available bandwidth being much greater than the system requirement, the actual available bandwidth between nodes approaches the link bandwidth, and the impact of deployment schemes on available bandwidth gradually weakens. Except for the Random scheme, the performance gap between other schemes and (w/o, IA) TAIA-MD and TAIA-MD gradually narrows to within $20\%$. Moreover, across diverse link bandwidths, TAIA-MD improves delay performance by about $5\%$ and convergence performance by approximately $50\%$ compared to (w/o, IA) TAIA-MD scheme, which demonstrates the effectiveness of the individual-adaptive mechanism, i.e., \textbf{Algorithm 3}.

\par
\begin{figure}[h]
\vspace{-0.5em}
\centering
\includegraphics[scale=0.35]{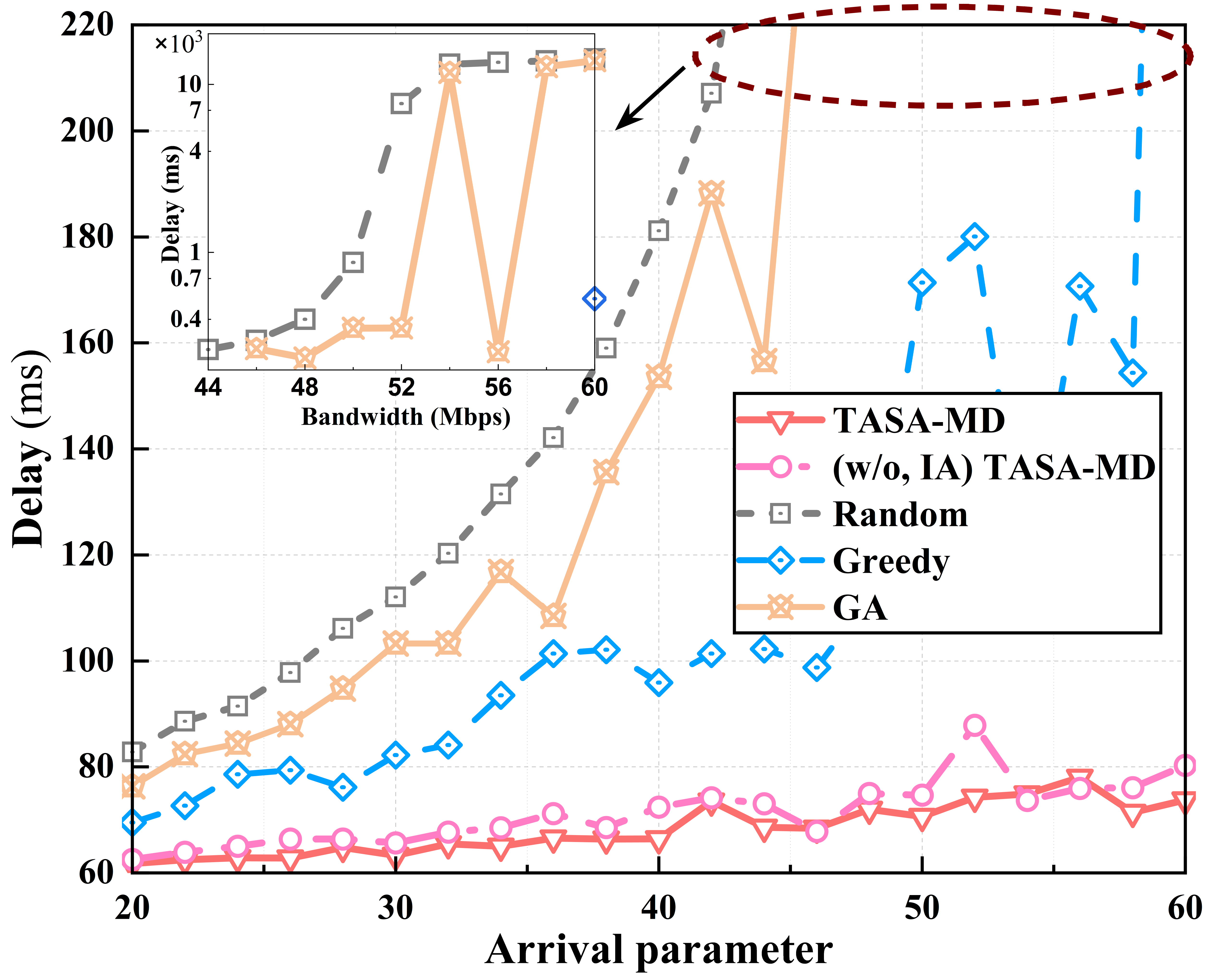}
\vspace{-0.6em}
\caption{\small Performance comparison between TAIA-MD and baseline schemes under different service arrival rates.}
\label{fig11}
\vspace{-0.2em}
\end{figure}

 Fig. \ref{fig11} illustrates the performance of five schemes across diverse service arrival rates. In scenarios with lower service arrival rates, denoting low-load conditions, except for random deployment, the performance differentials among these schemes appear marginal. However, (w/o, IA) TAIA-MD and TAIA-MD consistently exhibit the most superior performance. As service arrival rates increase, the bandwidth requirements of the system also increase accordingly. Compared with Random, Greedy, and GA schemes, the advantages of (w/o, IA) TAIA-MD and TAIA-MD become increasingly pronounced. This is primarily because they have taken into account the impact of microservice deployments on available bandwidth. As the service arrival rates increase, the burden borne by the links of edge networks intensifies, thereby leading to the realm of heightened network intricacies. Therefore, the proposed TAIA-MD scheme that integrates topology awareness and individual adaptability can fully utilize its advantages in these constantly changing networks.

\par
\begin{figure}[h]
\vspace{-0.5em}
\centering
\includegraphics[scale=0.35]{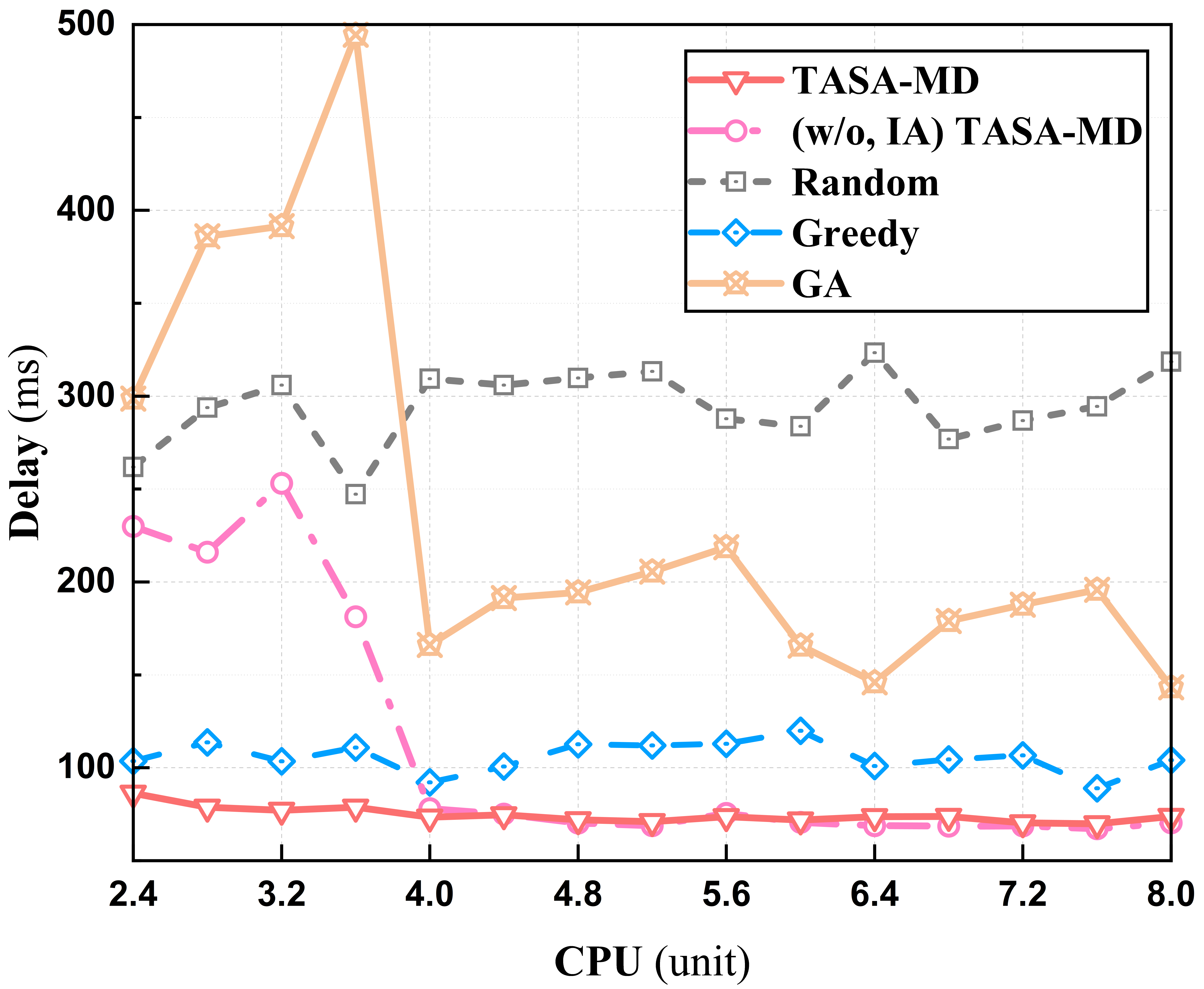}
\vspace{-0.6em}
\caption{\small Performance comparison between TAIA-MD and baseline schemes under different service arrival rates.}
\label{fig12}
\vspace{-0.8em}
\end{figure}

 Given sufficient memory resources, we adjust the computational resources on each node to evaluate the performance of algorithms. As shown in Fig. \ref{fig12}, with limited CPU resources, the flexibility of microservice deployment options is constrained, which makes the crossover and mutation process difficult for GA and (w/o, IA) TAIA-MD. In this case, Greedy and TAIA-MD, with individual adaptability, demonstrate superior performance. As computational resources increase, the flexibility of microservice deployment options expands, enhancing the performance of GA and (w/o, IA) TAIA-MD. When CPUs exceed four units, they surpass microservice deployment requirements, resulting in no significant performance changes for all schemes. This reveals that dynamically adjusting computational resources based on microservice deployment results can reduce resource overhead.

\par
\begin{figure}[h]
\vspace{-0.5em}
\centering
\includegraphics[scale=0.35]{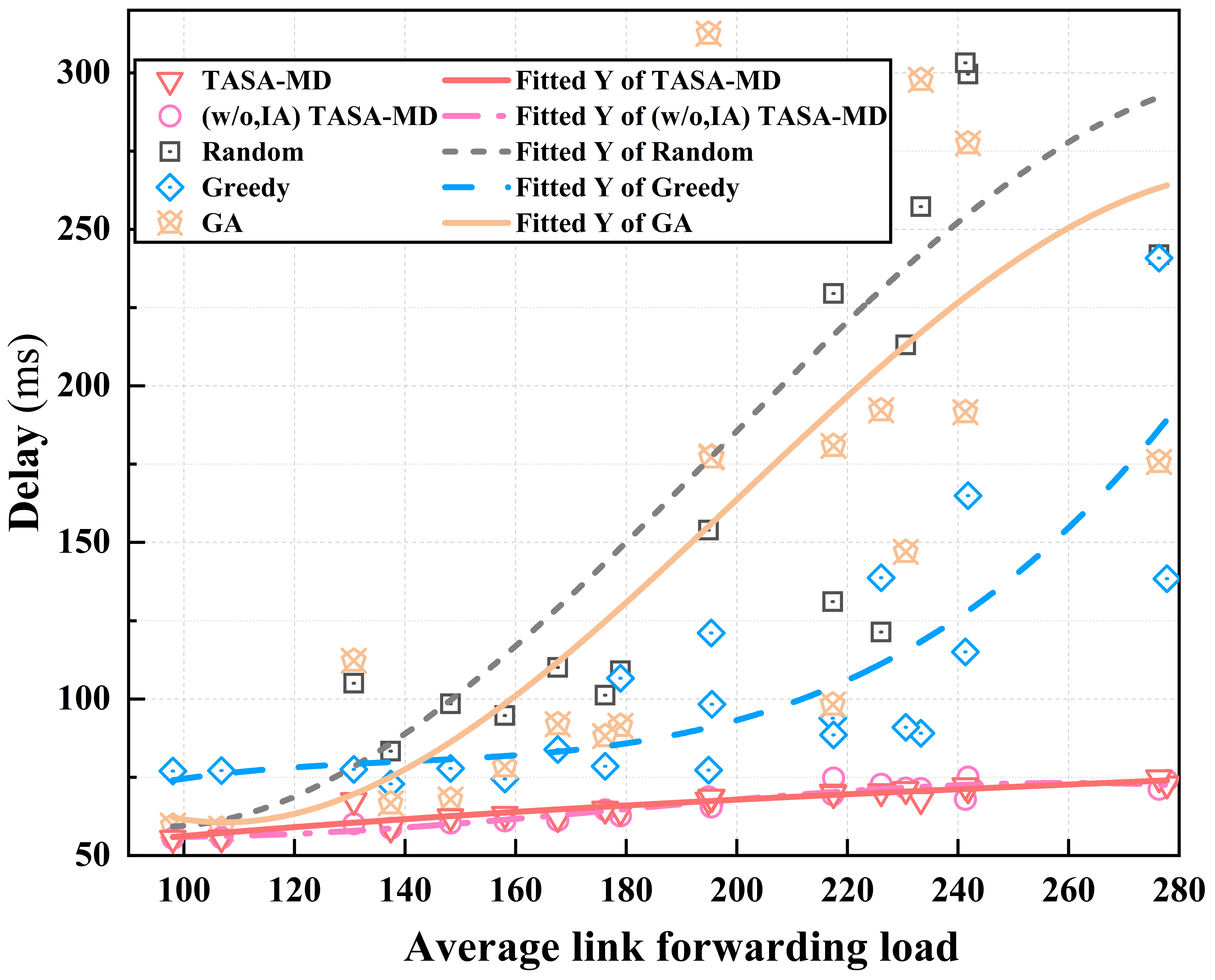}
\vspace{-0.6em}
\caption{\small Performance comparison between TAIA-MD and baseline schemes under different network topologies.}
\label{fig13}
\vspace{-0.2em}
\end{figure}

 To thoroughly investigate the impact of network topology on microservice deployment performance, we analyze the delay performance of TAIA-MD and baseline schemes under different average link forwarding loads, as shown in Fig. \ref{fig13}. As the average link forwarding load increases, the competition for bandwidth resources between microservices becomes more serious, leading to higher system delay. Under high average link forwarding loads, (w/o, IA) TAIA-MD and TAIA-MD exhibit significantly better delay performance compared to Random, GA, and Greedy, which demonstrates that (w/o, IA) TAIA-MD and TAIA-MD effectively address bandwidth collision in edge networks. Furthermore, although (w/o, IA) TAIA-MD and TAIA-MD achieve similar delay performance, Fig. \ref{fig9} has already demonstrated TAIA-MD's superior convergence performance.

\begin{figure*}[t]
\centering
  \subfloat[]{
   \includegraphics[scale=0.36]{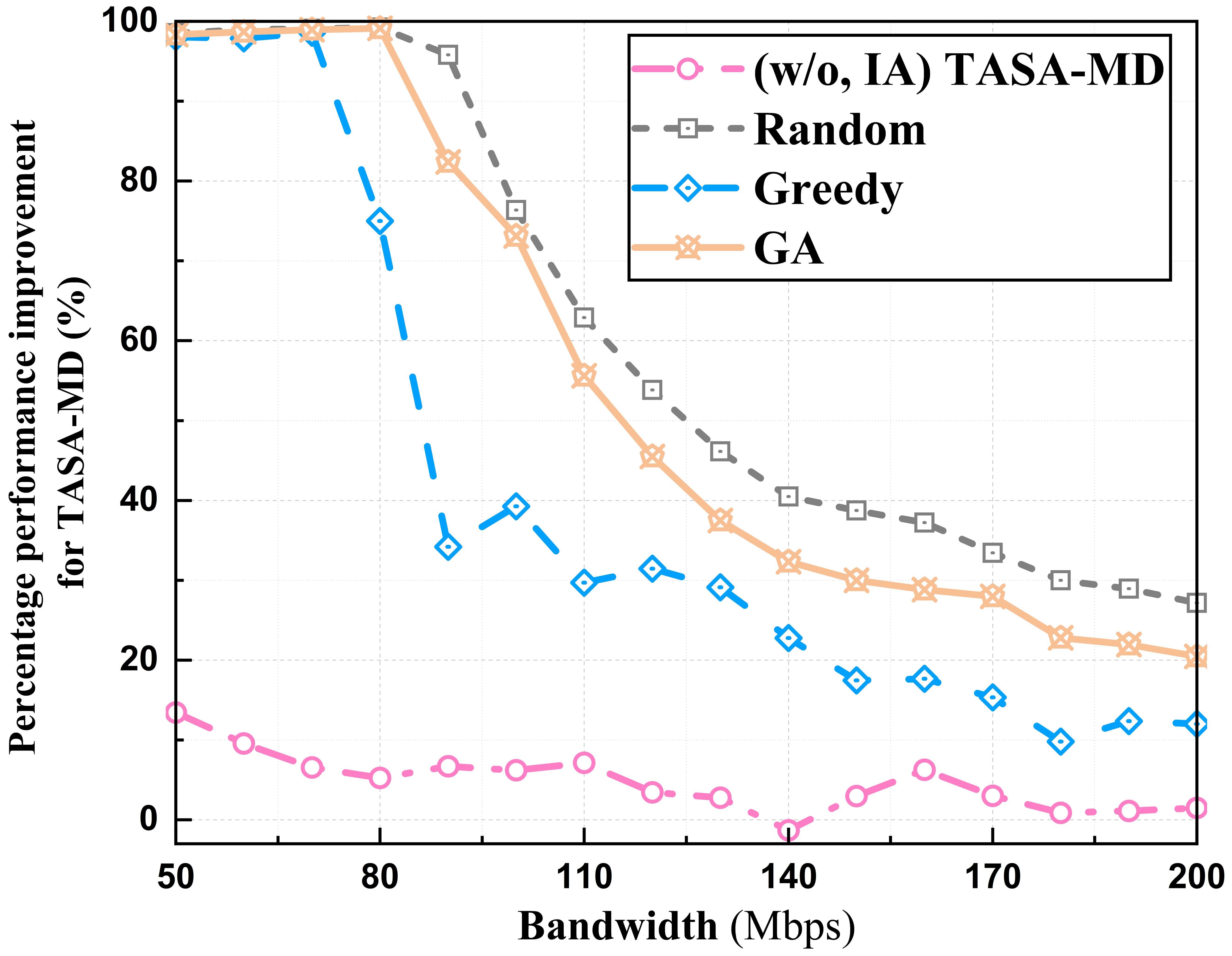}
  }
  \hspace{0.5cm} 
  \subfloat[]{
   \includegraphics[scale=0.36]{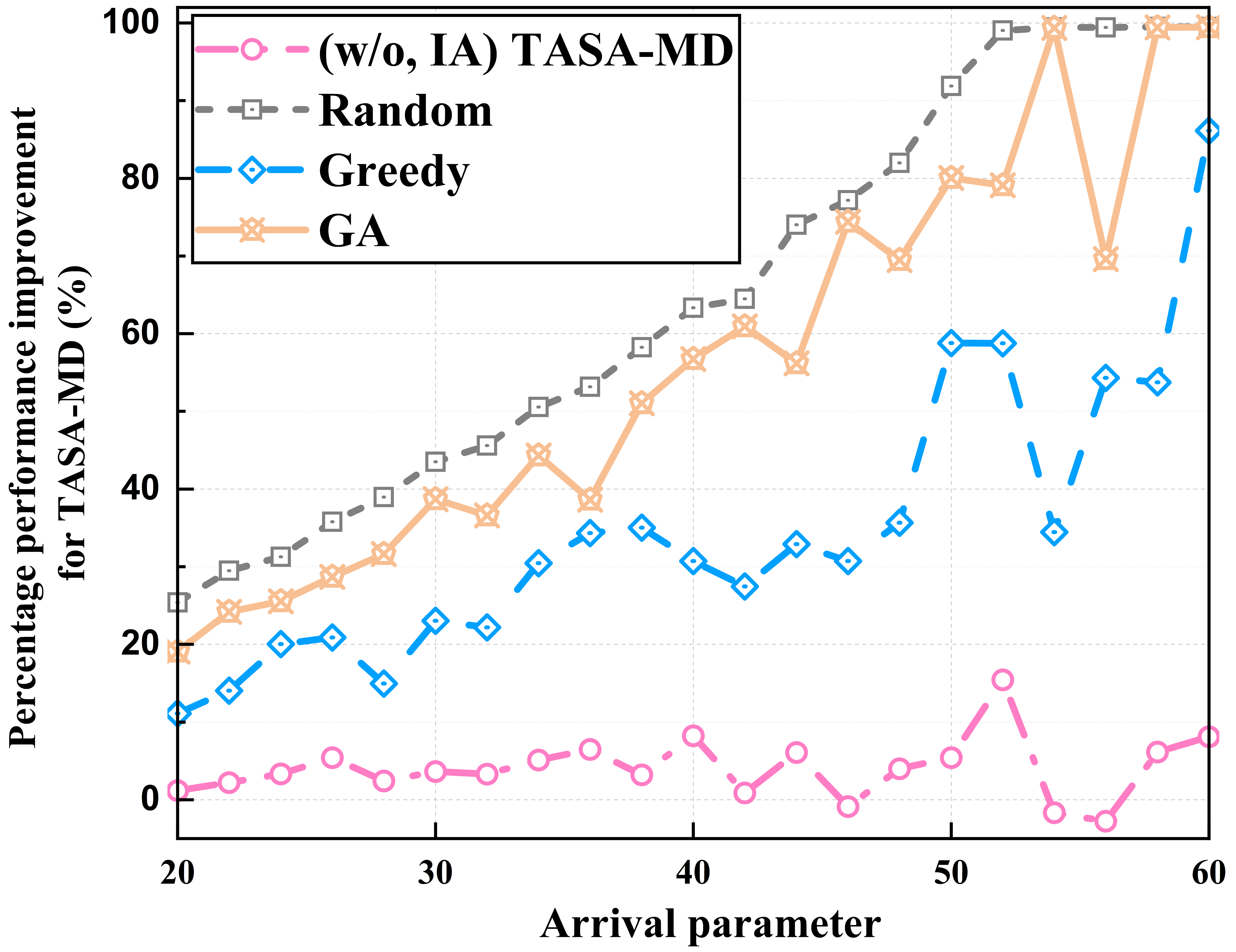}
  }\\[0.5cm] 
  \subfloat[]{
   \includegraphics[scale=0.36]{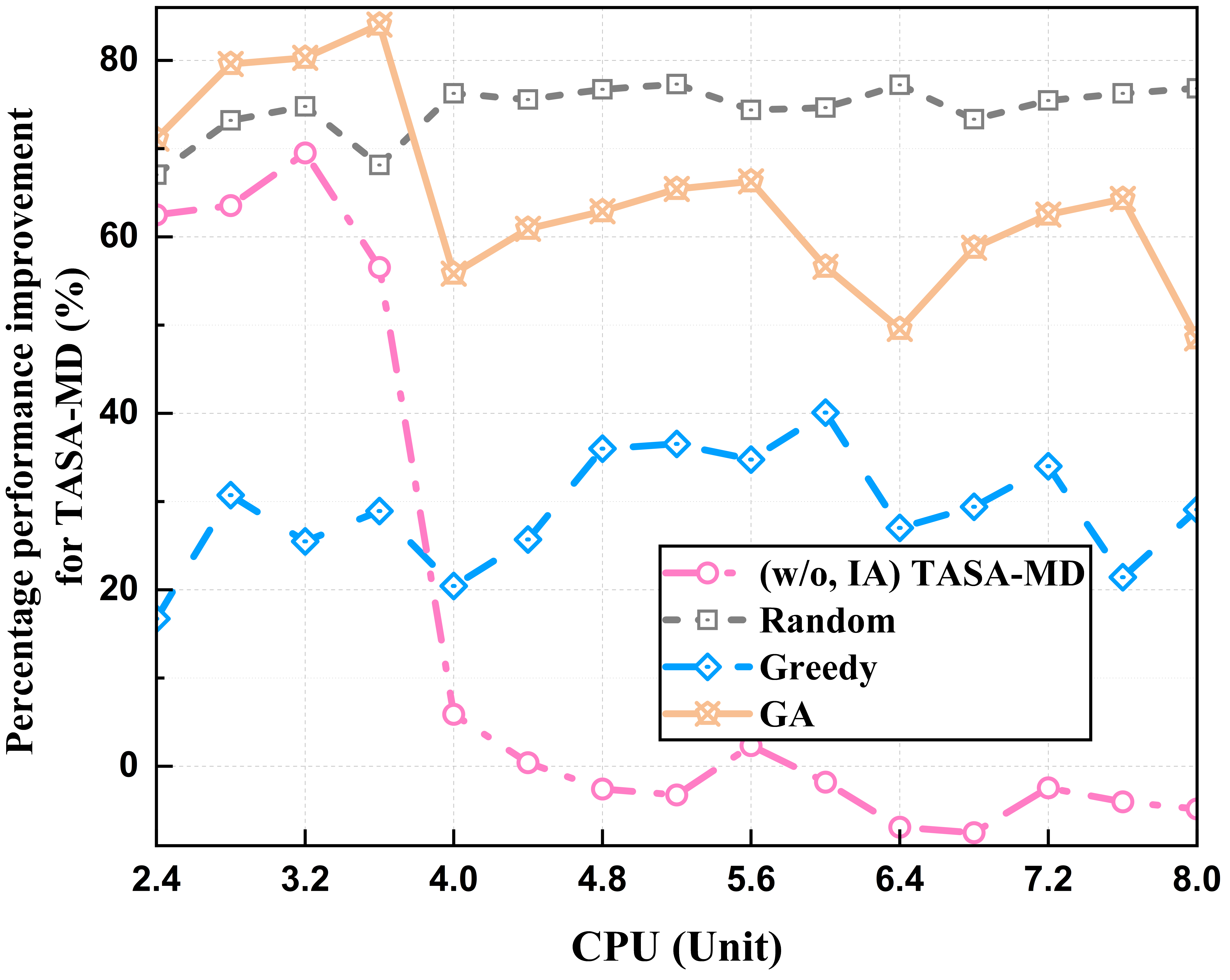}
  }
  \hspace{0.5cm} 
  \subfloat[]{
   \includegraphics[scale=0.36]{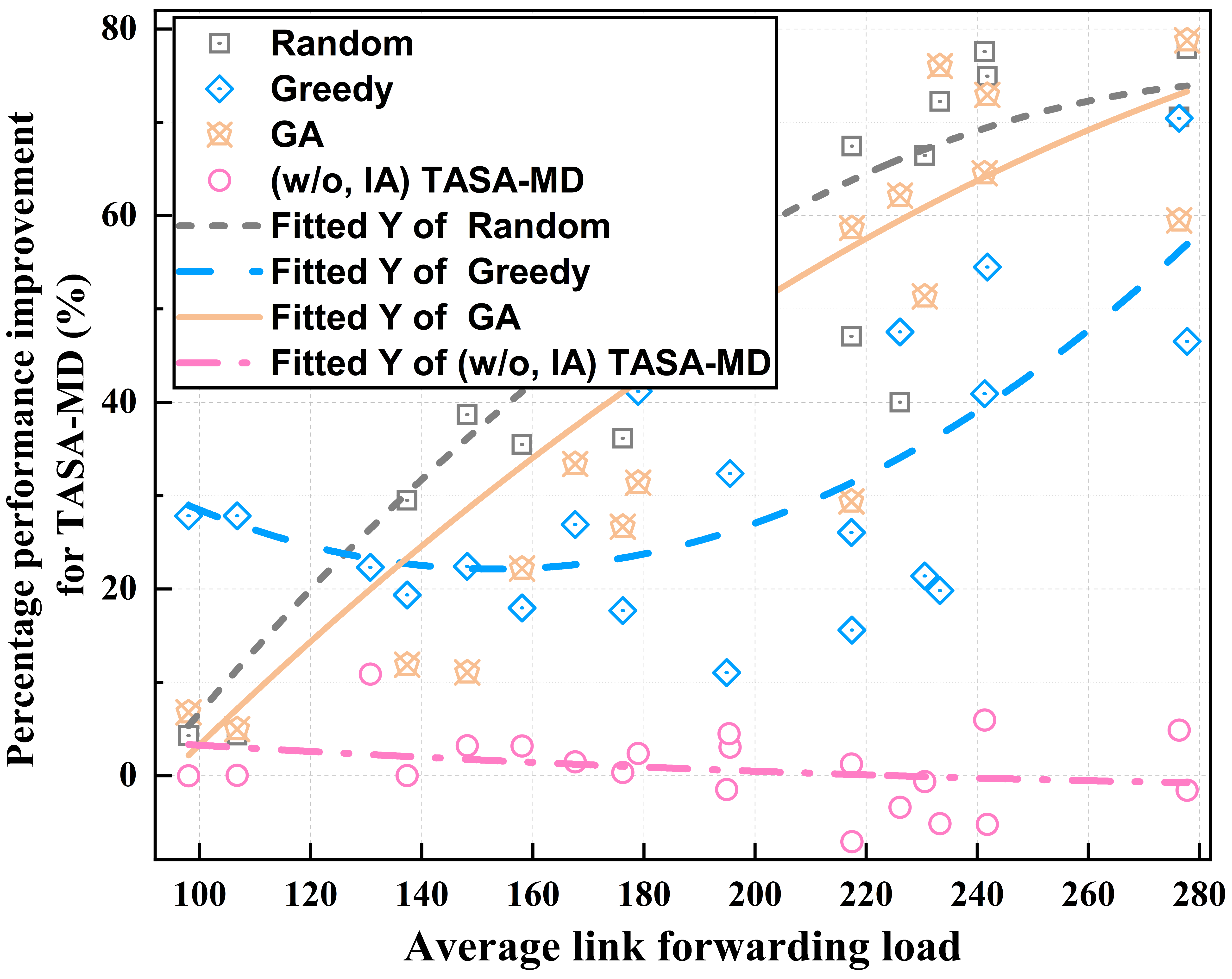}
  }
  \caption{\small Performance improvement percentage between TAIA-MD and baseline schemes under different (a) bandwidths, (b) service arrival rates, (c) computational resources, and (d) network topologies.}
  \label{fig14}
  \vspace{-1.3em}
\end{figure*}

\par Assume the delays for Scheme A and Scheme B are $T_{A}$ and $T_{B}$, respectively. Then, the performance improvement percentage of Scheme A over Scheme B can be calculated as $\frac{T_{B}-T_{A}}{T_{B}} \times 100\%$. As shown in Fig. \ref{fig14}, we present the performance improvement percentage of TAIA-MD compared to baseline schemes under varied bandwidths, service arrival rates, computational resources, and average link forwarding loads. In most cases, considering the effects of topology and traffic on available bandwidth during deployment scheme calculation can effectively reduce the system's delay. TAIA-MD can decrease system delay by approximately $30\%$-$60\%$ compared to other baseline schemes assuming fixed link bandwidth. Remarkably, in scenarios with low bandwidth, high service arrival rates, and complex topologies, TAIA-MD significantly prevents link congestion, providing substantial performance improvements. Furthermore, as illustrated in Fig. \ref{fig14} (c), TAIA-MD generally suggests a performance gain of about $5\%$ over (w/o, IA) TAIA-MD. Importantly, with limited CPU resources, TAIA-MD performs markedly better than (w/o, IA) TAIA-MD, which underscores TAIA-MD's extraordinary performance with topology-aware and individual-adaptive mechanisms for edge networks.

\vspace{-1em}

\subsection{DMSA Platform Validation Results}
\par
\begin{figure}[h]
    \centering
    \begin{minipage}[b]{0.48\textwidth}
        \centering
        \begin{subfigure}{0.48\columnwidth}
            \centering
            \includegraphics[width=\linewidth]{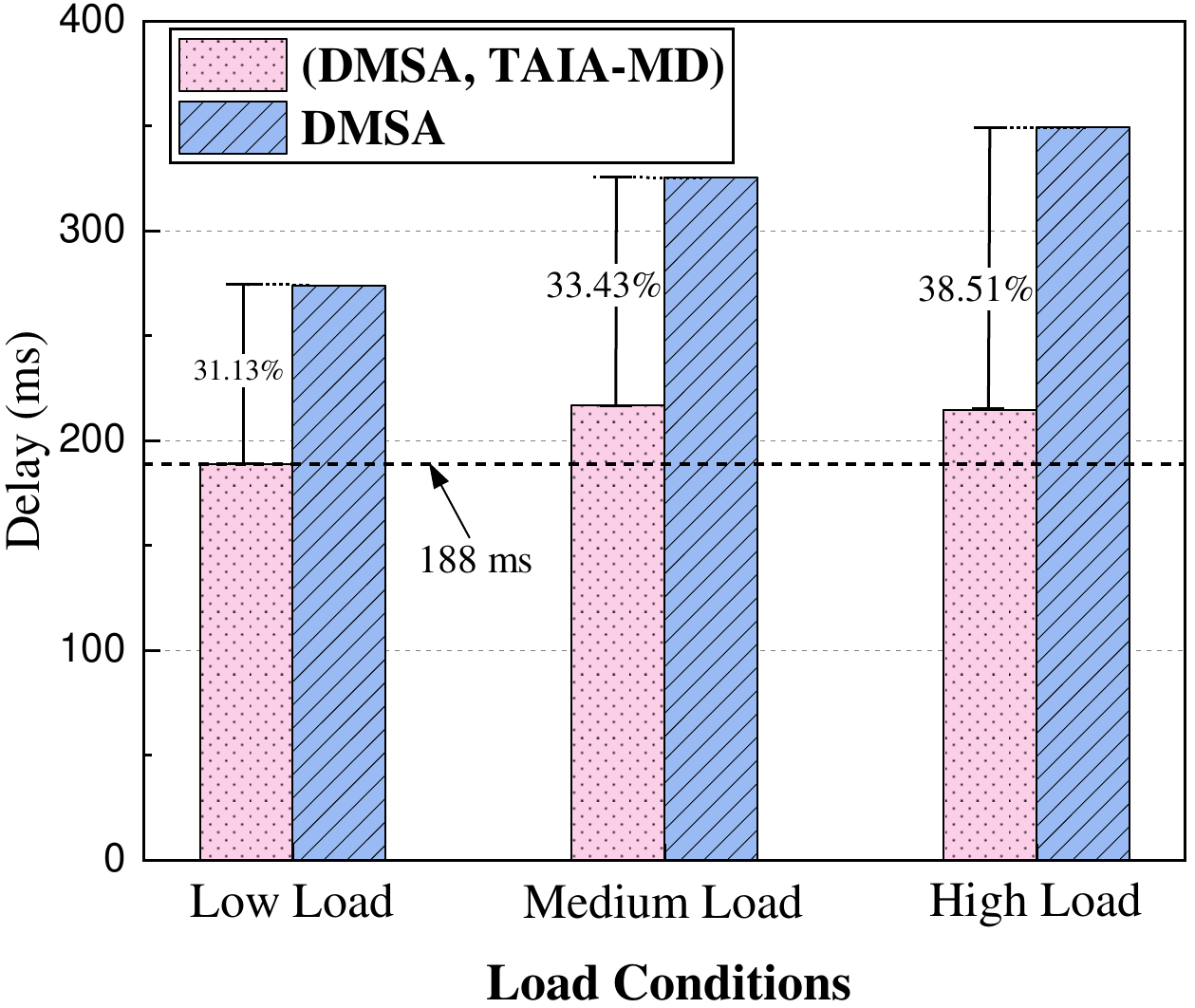}
            \caption{Graphic-text Services}
        \end{subfigure}
        \hfill
        \begin{subfigure}{0.48\columnwidth}
            \centering
            \includegraphics[width=\linewidth]{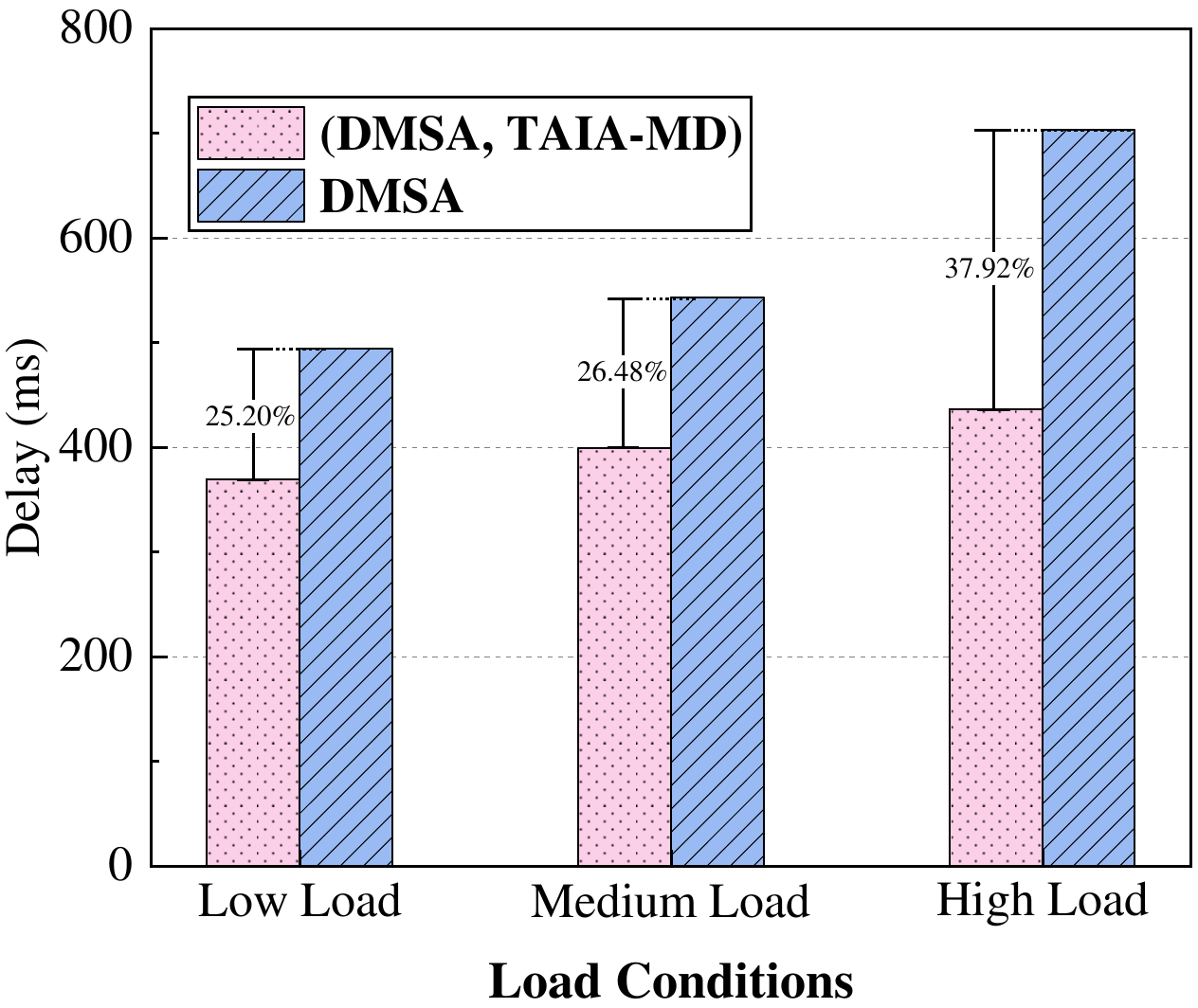}
            \caption{Video Services}
        \end{subfigure}
        \vspace{-0.3em}
        \caption{\small The average response delay of the TAIA-MD scheme for graphic-text and video services across different load conditions on the DMSA platform.}
        \vspace{0.3em}
        \label{fig15}
    \end{minipage}
    \vspace{-0.5em}
\end{figure}

 For latency-sensitive services like video and graphic-text services, we focus on analyzing the average delay performance of the TAIA-MD scheme deployed on the DMSA platform. As shown in Fig. \ref{fig15}, the average delay under various load conditions reveals that the DMSA platform with the TAIA-MD scheme (i.e., (DMSA, TAIA-MD) scheme) significantly outperforms the standalone DMSA platform. Specifically, for graphic-text services, the average delay improvement of the (DMSA, TAIA-MD) scheme over standalone the DMSA scheme is substantial, ranging from 31.13\% under low loads to 38.51\% under high loads. For video services, the performance enhancement of average delay for the (DMSA, TAIA-MD) scheme under low, medium, and high loads is 25.20\%, 26.48\%, and 37.92\%, respectively. It can be seen that as the load increases, the delay performance advantage of the (DMSA, TAIA-MD) scheme becomes increasingly significant. In addition, the average delay of the DMSA platform deteriorates with the increase of load for both graphic-text and video services, while the average delay of the (DMSA, TAIA-MD) scheme is less affected by increasing load. This is primarily attributed to microservice deployment optimization of the (DMSA, TAIA-MD) scheme, which significantly reduces the overhead of instance scheduling in the DMSA platform and thus achieves lower average delay.

\par
\begin{figure*}[t]
    \centering
    \begin{subfigure}{0.6\columnwidth}
        \centering
        \includegraphics[width=\linewidth]{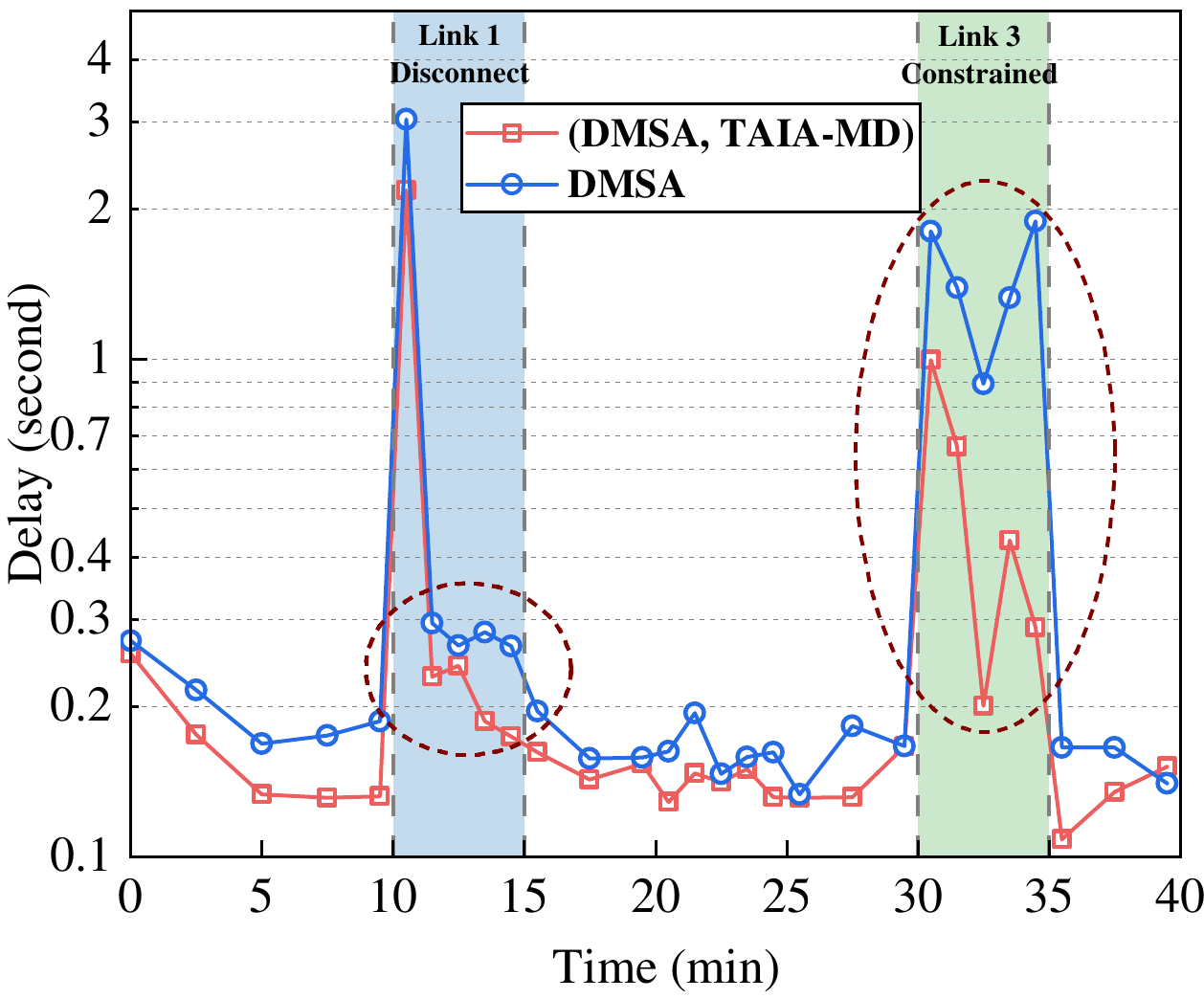}
        \caption{\small Low Load}
    \end{subfigure}
    \hspace{1em}
    \begin{subfigure}{0.6\columnwidth}
        \centering
        \includegraphics[width=\linewidth]{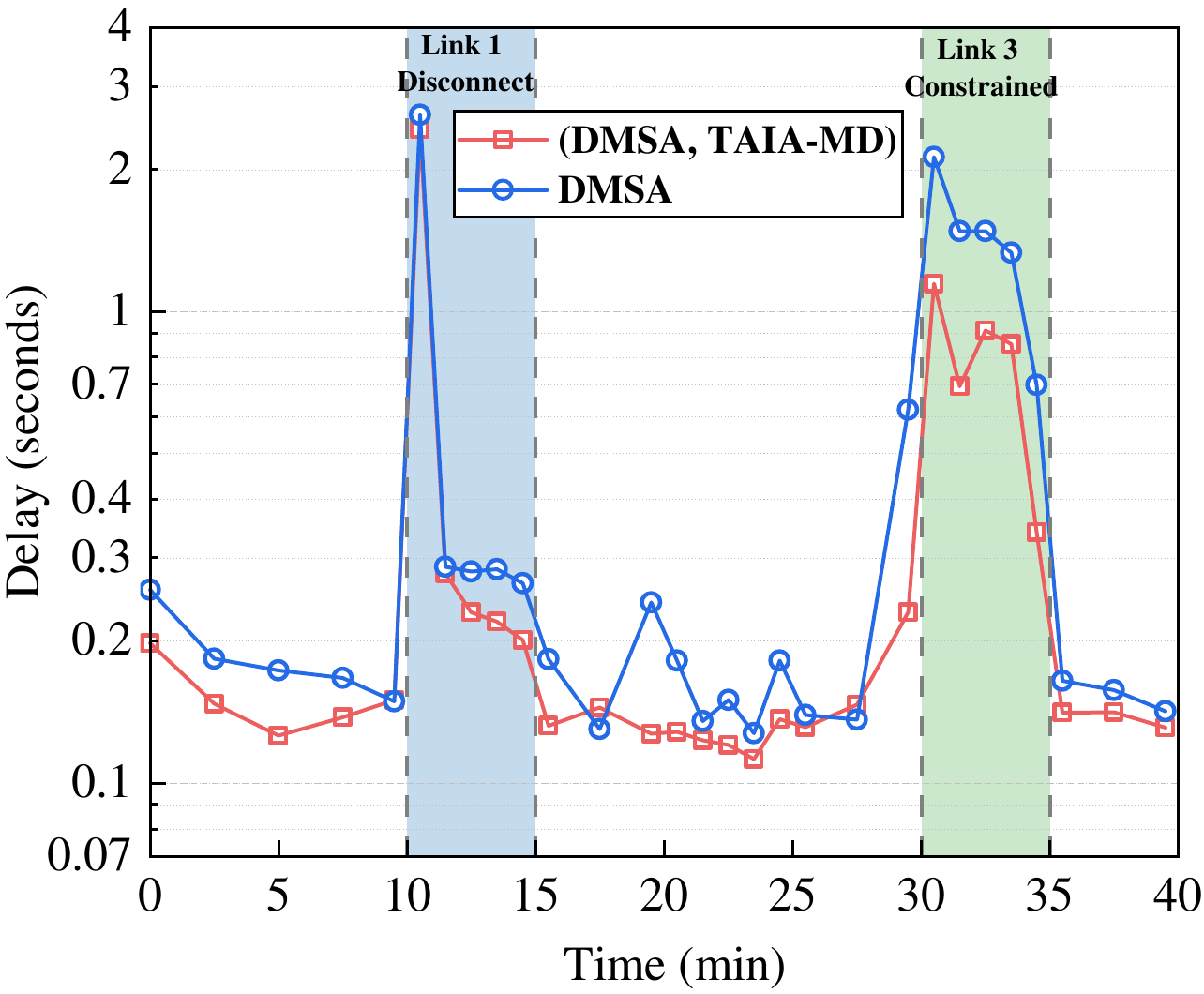}
        \caption{\small Medium Load}
    \end{subfigure}
    \hspace{1em}
    \begin{subfigure}{0.6\columnwidth}
        \centering
        \includegraphics[width=\linewidth]{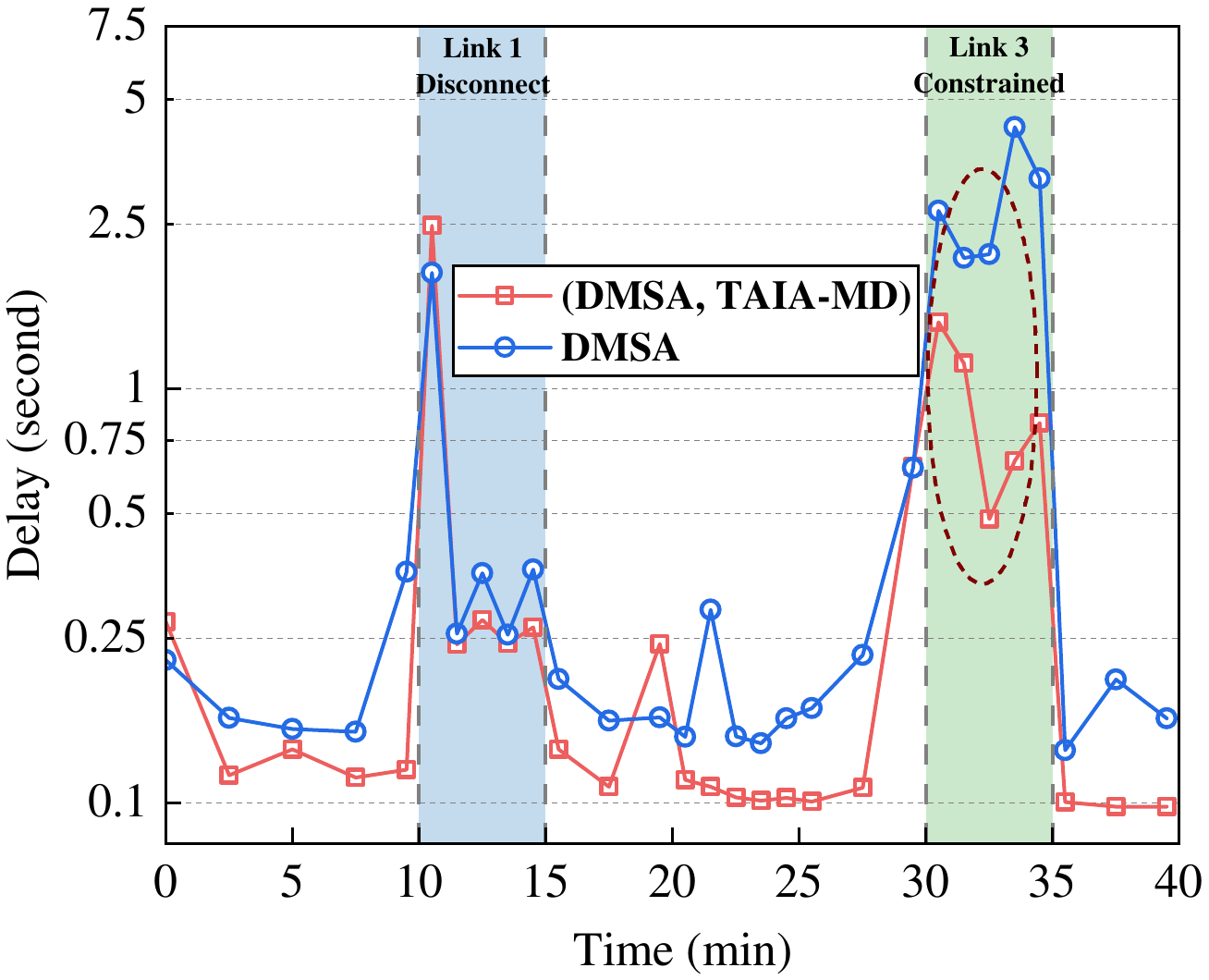}
        \caption{\small High Load}
    \end{subfigure}
    \caption{\small Dynamic response delay of video services under different load conditions on the DMSA Platform with the TAIA-MD Scheme.}
    \label{fig16}
    \vspace{-0.5em}
\end{figure*}

 As illustrated in Fig. \ref{fig16}, we present the dynamic variations in delay performance over run time under various load conditions, taking video services as an example. Fig. \ref{fig16} (a), (b), and (c) demonstrate that the (DMSA, TAIA-MD) scheme consistently outperforms the standalone DMSA platform in terms of delay performance under different loads. In particular, during the $10 \sim 15$ \emph{min} interval, a link disruption occurs between Switch 1 and Node 3, which causes a sharp increase in delay for both the (DMSA, TAIA-MD) scheme and DMSA. However, the delay performance can quickly recover from the link disconnection. Remarkably, the (DMSA, TAIA-MD) scheme restores network performance more rapidly. This is primarily due to the fact that the TAIA-MD scheme optimizes instance deployment on the basis of the DMSA platform's effective sensing of network load and edge node status, and dynamically adjusts the scheduling strategies. During the $30 \sim 35$ \emph{min} interval, the link bandwidth between Switch 3 and Switch 5 fluctuates and drops sharply to $100$ Mbps. During this period, user requests are still routed through the constrained link, resulting in a significant increase in delay for both the (DMSA, TAIA-MD) scheme and DMSA platform. However, the delay performance can recover shortly after. The (DMSA, TAIA-MD) scheme is less affected by network fluctuations and recovers faster. This demonstrates the enhanced robustness and adaptability of the DMSA platform with the TAIA-MD scheme in handling network emergencies.

\vspace{-0.2em}

\par
\begin{figure*}[t]
    \centering
    \begin{subfigure}{0.6\columnwidth}
        \centering
        \includegraphics[width=\linewidth]{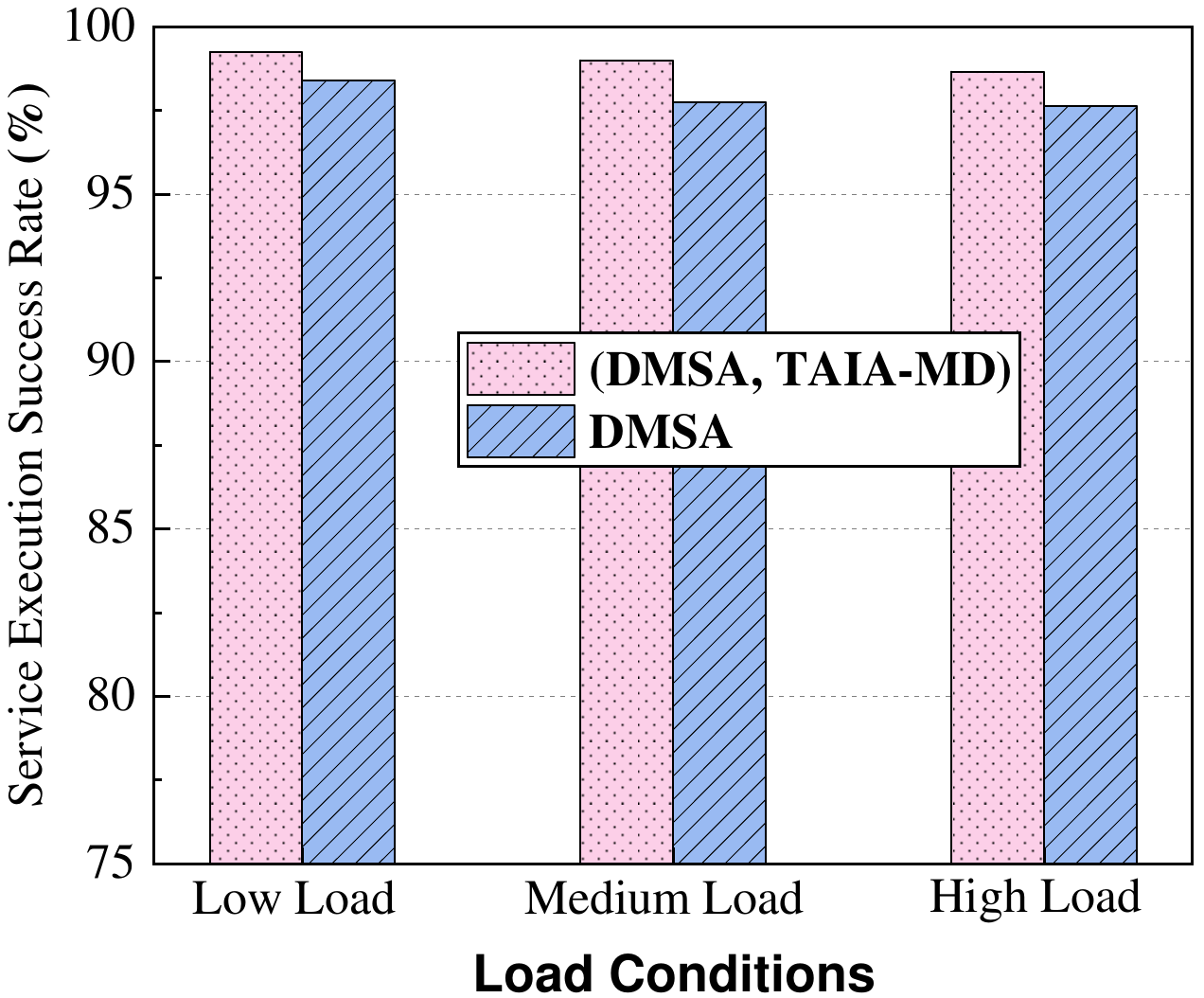}
        \caption{\small Graphic-text Services}
    \end{subfigure}
    \hspace{1em}
    \begin{subfigure}{0.6\columnwidth}
        \centering
        \includegraphics[width=\linewidth]{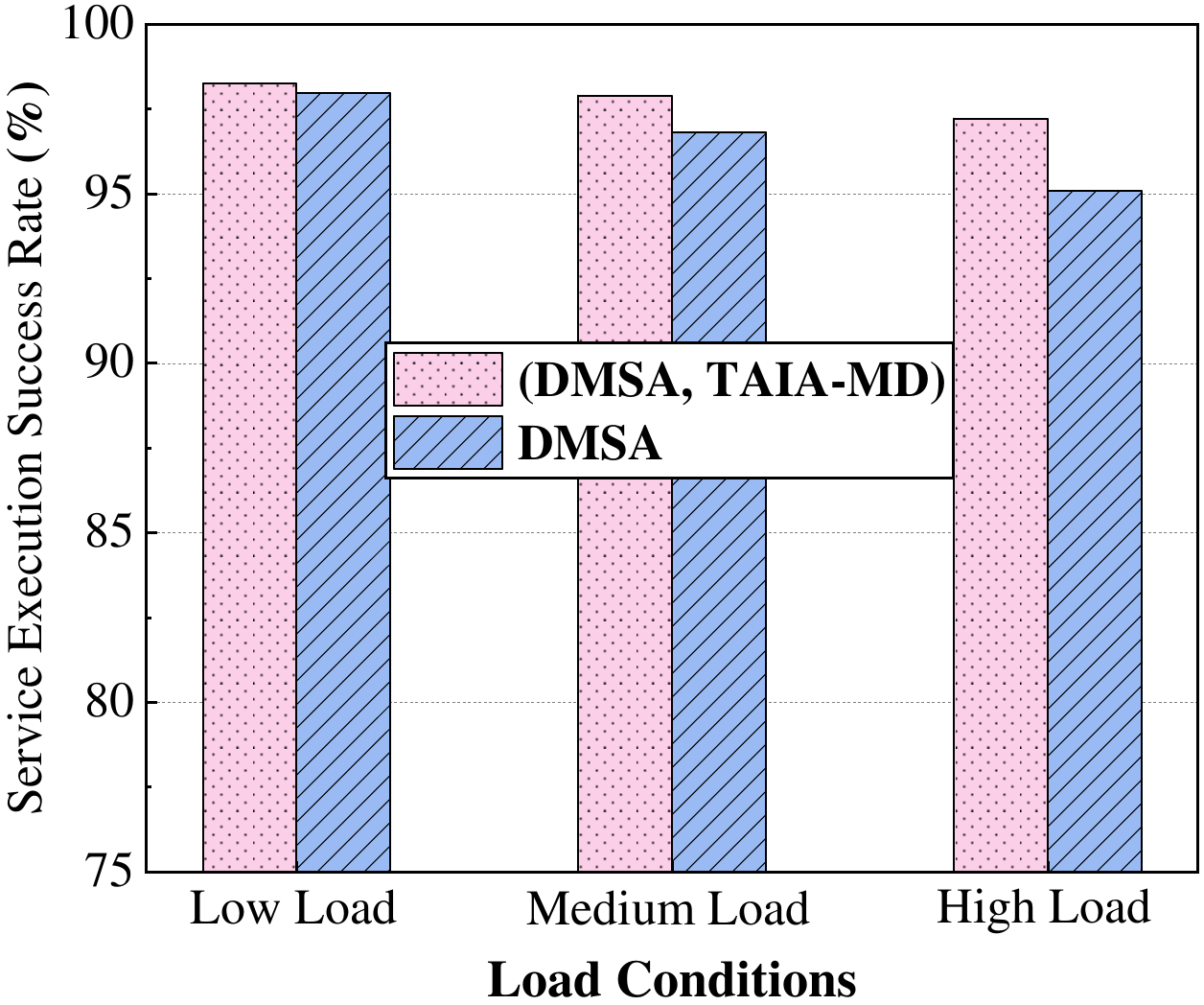}
        \caption{\small Video Services}
    \end{subfigure}
    \hspace{1em}
    \begin{subfigure}{0.6\columnwidth}
        \centering
        \includegraphics[width=\linewidth]{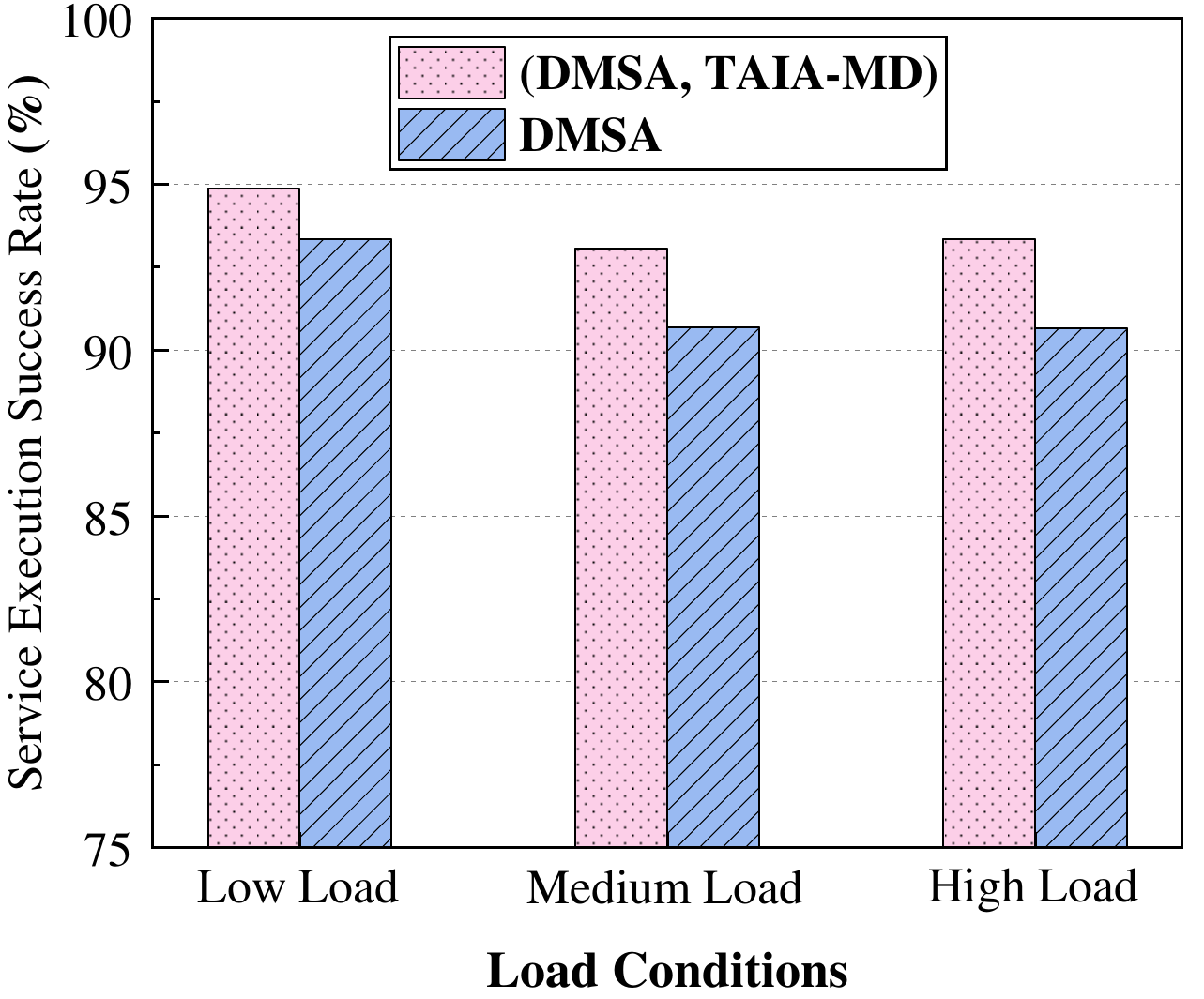}
        \caption{\small File Download Services}
    \end{subfigure}
    \caption{\small The TAIA-MD scheme on the service execution success rate under different services on the DMSA platform.}
    \label{fig17}
    \vspace{-1em}
\end{figure*}

 Next, the robustness of the (DMSA, TAIA-MD) scheme and DMSA platform is analyzed. As shown in Fig. \ref{fig17} (a), (b), and (c), we examine the average service execution success rates of graphic-text, video, and file download services under various load conditions. Compared to the performance of the DMSA platform itself, the (DMSA, TAIA-MD) scheme overall improves the service execution success rates in edge networks and significantly enhances their robust performance. Fig. \ref{fig17} (a)-(c) demonstrate that the (DMSA, TAIA-MD) scheme exhibits more pronounced performance advantages in microservice deployments as service data volume and load increase. For graphic-text and video services, the (DMSA, TAIA-MD) scheme stably maintains the service execution success rates of more than 98\% and 97\%, respectively. In large file downloads, the delay performance improvements in service execution success rates with the (DMSA, TAIA-MD) scheme become notably more significant as the load increases compared to the DMSA platform alone.

\section{CONCLUSION AND FUTURE OUTLOOK}

\vspace{-0.3em}

\par In this paper, we have introduced an innovative microservice deployment architecture, which integrates a three-tier network traffic model encompassing the service layer, microservices layer, and edge node layer. This traffic model meticulously characterizes the complex dependencies between edge network topology and microservices, and maps microservice deployments onto link traffic to accurately estimate communication delay. On this basis, we have formulated a weighted sum communication delay optimization problem to minimize communication delay by optimizing microservices deployments. To effectively solve this problem, we have proposed a novel deployment scheme called TAIA-MD, which accurately senses the network topology and incorporates an individual-adaptive mechanism in GA to accelerate convergence and avoid local optima. Extensive simulations show that in bandwidth-constrained edge networks with complex topologies, TAIA-MD improves the delay performance by approximately 30\% to 60\% compared to existing deployment schemes. Moreover, through real-world deployments on the DMSA platform, we have demonstrated the robustness of the TAIA-MD scheme for withstanding link failures and network fluctuations and validated its practicality in MSA for edge networks.

\par There are still limitations to this study. In particular, in the analysis of the microservice deployment problem, we focused on static routing schemes in edge networks. However, there actually may be multiple reachable paths between nodes, and alterations in routing schemes might impact link loads, consequently influencing the communication performance of services. Therefore, in future work, we intend to explore dynamic routing schemes to develop more comprehensive link traffic analysis models. In addition, we will also consider routing strategies as variable factors for joint optimization of routing strategies and microservice deployment schemes.

\vspace{-0.5em}

\footnotesize
\bibliographystyle{IEEEtran}
\bibliography{IEEEabrv,ref}
\vspace{-0.8cm}
\begin{IEEEbiography}[{\includegraphics[width=1in,height=1.25in,clip,keepaspectratio]{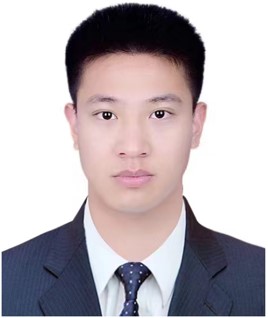}}]{Yuang Chen (Graduate Student Member, IEEE)}
	received the B.S. degree from the Hefei University of Technology (HFUT), Hefei, China, in 2021. He is currently pursuing the master's degree with the Department of Electronic Engineering and Information Science, University of Science and Technology of China (USTC), Hefei, China. His research interests include 5G/6G wireless network technologies, particularly analyzing key performance indicators such as reliability, latency, jitter, and AoI of URLLLC services using stochastic network calculus theory, extreme value theory, and large deviation theory.
\end{IEEEbiography}

\begin{IEEEbiography}[{\includegraphics[width=1in,height=1.25in,clip,keepaspectratio]{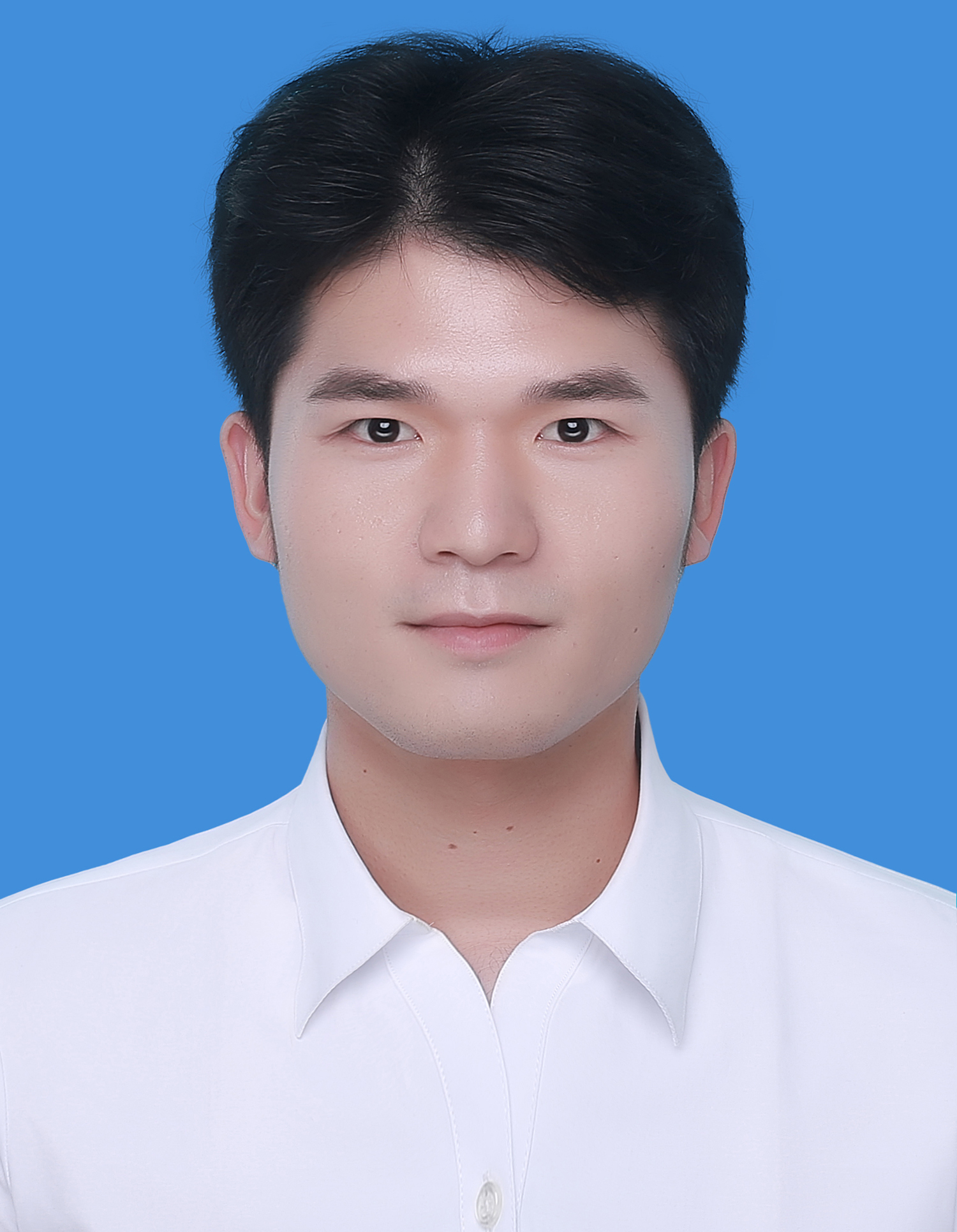}}]{Chang Wu}
	received the B.S. degree from Dalian Maritime University (DLMU), Dalian, China, in 2021. He is currently pursuing the Ph.D. degree in communication and information systems with the Department of Electronic Engineering and Information Science, University of Science and Technology of China (USTC), Hefei, China. His research interests include 5G/6G wireless network technologies such as traffic engineering, radio resource management and multiple access networks.
\end{IEEEbiography}

\begin{IEEEbiography}[{\includegraphics[width=1in,height=1.25in,clip,keepaspectratio]{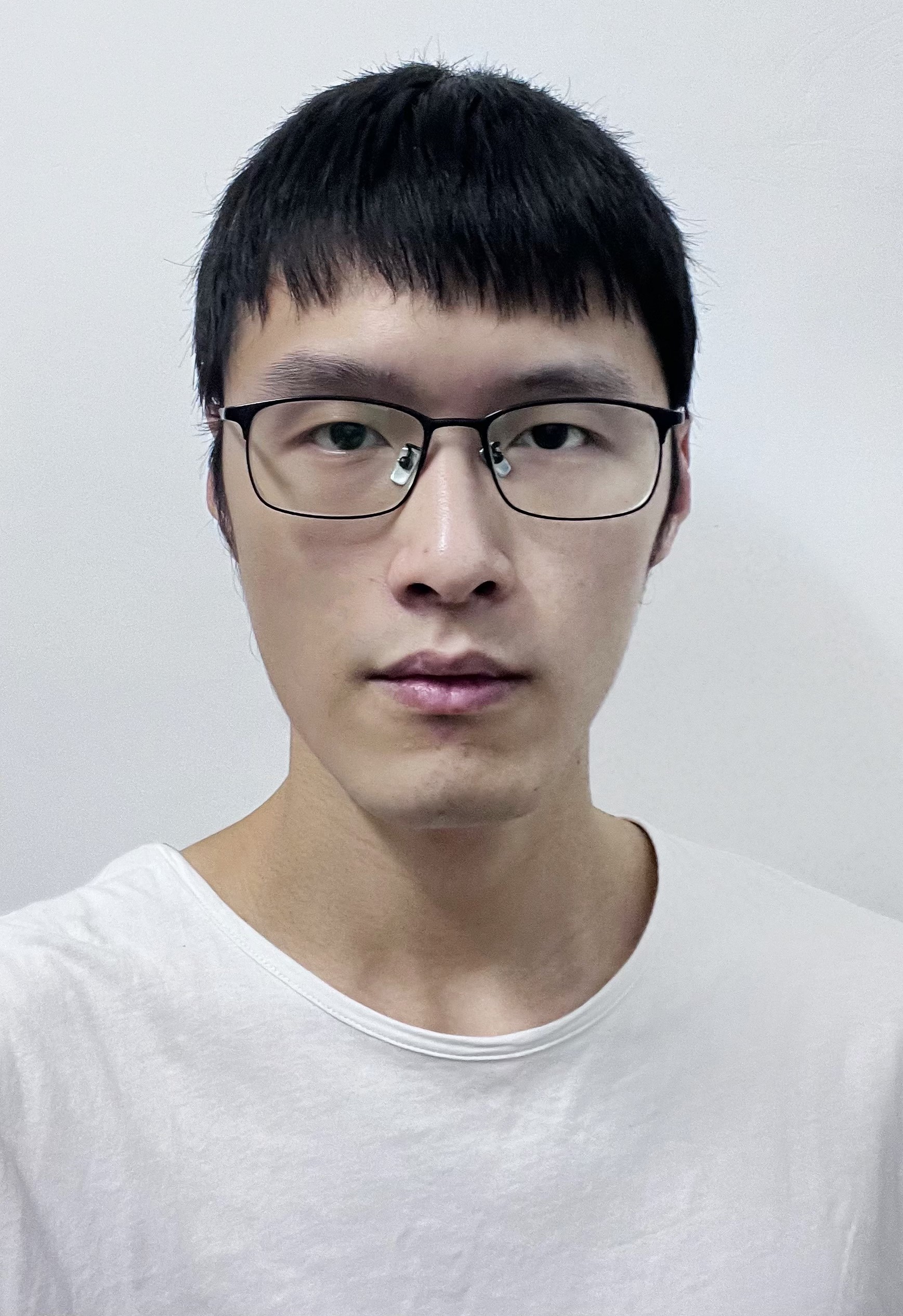}}]{Fangyu Zhang}
	received the B.S. degree from the School of University of Science and Technology of China (USTC), Hefei, China, in 2020. He is currently pursuing the Ph.D. degree with the Department of Electronic Engineering and Information Science, USTC, Hefei, China. His research interests include machine learning, network function virtualization, and network resource allocation in heterogeneous networks.
\end{IEEEbiography}

\vspace{-1cm}

\begin{IEEEbiography}[{\includegraphics[width=1in,height=1.25in,clip,keepaspectratio]{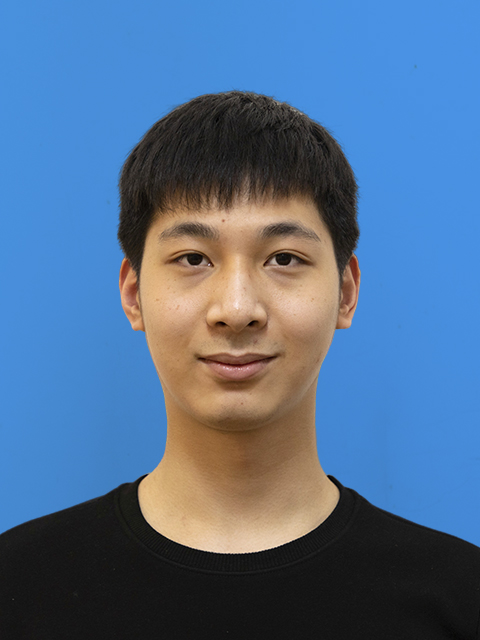}}]{Chengdi Lu}
	received the BS degree from the Department of Electronic Engineering and Information Science, USTC in July, 2020. He is currently working toward the PhD degree with the Department of Electronic Engineering and Information Science. His research interests include data center networks and programmable switches.
\end{IEEEbiography}

\vspace{-1cm}

\begin{IEEEbiography}[{\includegraphics[width=1in,height=1.25in,clip,keepaspectratio]{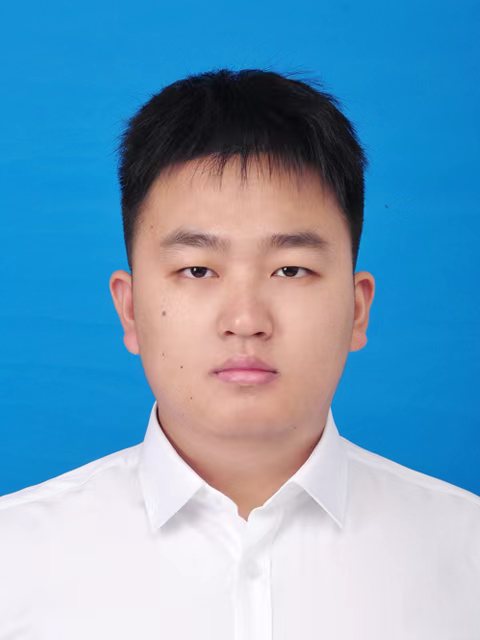}}]{Yongsheng Huang}
	received the B.S. degree in information engineering from University of Electronic Science and Technology of China (UESTC), Chengdu, China, in 2021. He received the master's degree with the Department of Electronic Engineering and Information Science, University of Scienceand Technology of China (USTC), Hefei, China. His research interests include microservice.
\end{IEEEbiography}

\vspace{-1cm}

\begin{IEEEbiography}[{\includegraphics[width=1in,height=1.25in,clip,keepaspectratio]{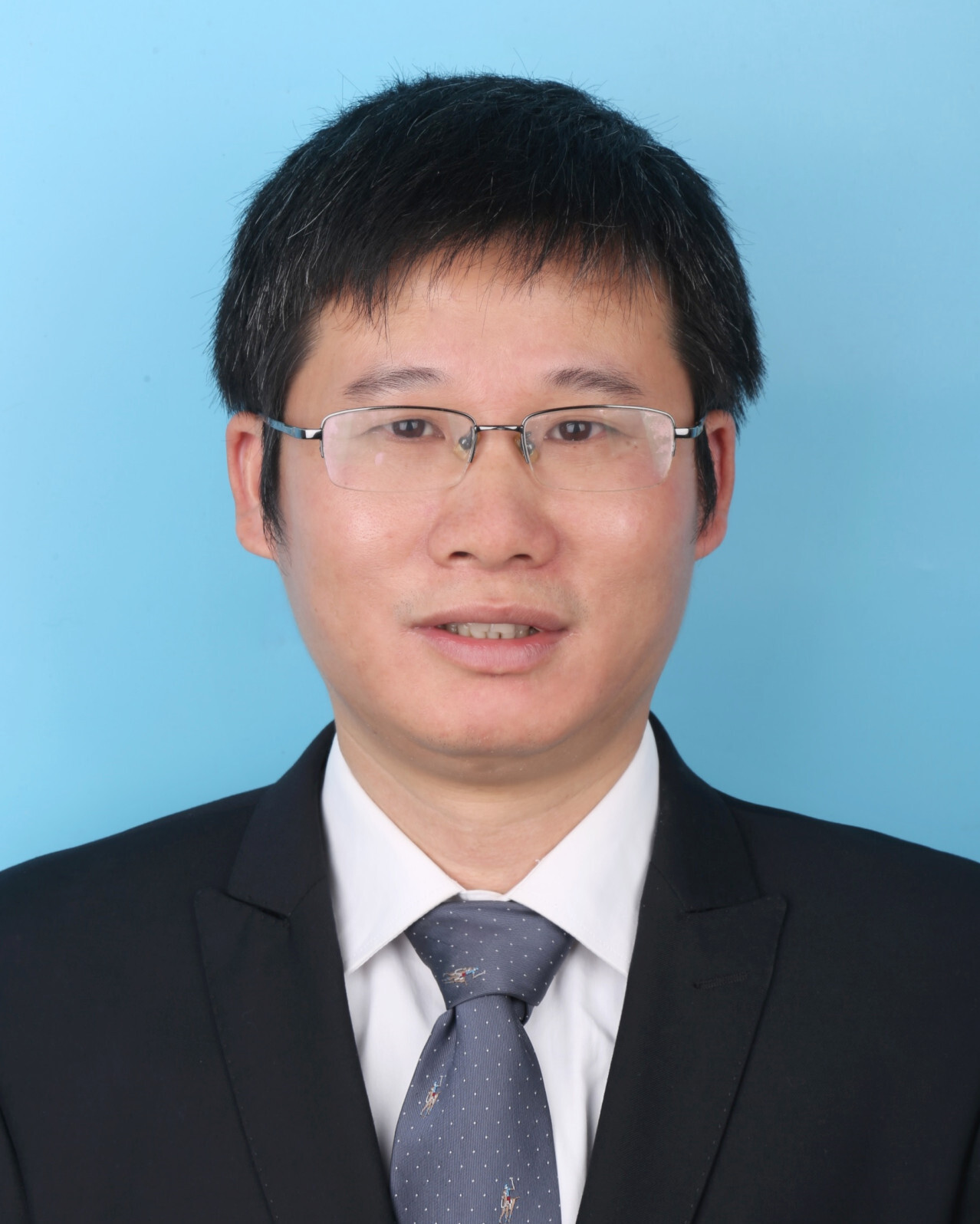}}]{Hancheng Lu (Senior Member, IEEE)}
	received his Ph.D. in communication and information systems from the University of Science and Technology of China, Hefei, China, in 2005. He is currently a tenured professor in the Department of Electronic Engineering and Information Science at the University of Science and Technology of China. He is also working at the Hefei National Comprehensive Science Center Artificial Intelligence Research Institute located in Hefei, China. He has rich research experience in multimedia communication, wireless edge networks, future network architecture and protocols, as well as machine learning algorithms for network communication, involving scheduling, resource management, routing, transmission, and other fields. In the past 5 years, more than 80 papers have been published in top journals such as IEEE Trans and flagship conferences such as IEEE INFOCOM in this field, and have won the Best Paper Award of IEEE GLOBECOM 2021 and the Best Paper Award of WCSP 2019 and WCSP 2016 in the field of communication. In addition, he currently serves as an editorial board member for numerous journals such as IEEE Internet of Things Journal, China Communications, and IET Communications.
\end{IEEEbiography}

\vspace{-1cm}
\end{document}